\documentclass[a4paper,fleqn,usenatbib]{mnras}

\usepackage[T1]{fontenc}
\usepackage{ae,aecompl}

\usepackage{graphicx}	
\usepackage{amsmath}	
\usepackage{amssymb}	

\title[Flare Induced Global Waves in the Sun]{On the signatures of flare-induced global waves 
in the Sun: GOLF and VIRGO observations}

\author[Brajesh Kumar et al.]{
Brajesh Kumar,$^{1}$\thanks{E-mail: brajesh@prl.res.in (BK)}
Savita Mathur,$^{2}$
Rafael A. Garc\'ia$^{3}$
and Antonio Jim\'enez$^{4,5}$
\\
$^{1}$Udaipur Solar Observatory, Physical Research Laboratory, Dewali, Badi Road, Udaipur 313004, Rajasthan, India \\
$^{2}$Space Science Institute, 4750 Walnut Street, Suite 205, Boulder, CO 80301, USA \\
$^{3}$Laboratoire AIM, CEA/DRF-CNRS, Universit\'e Paris 7 Diderot, IRFU/SAp, Center de Saclay, 91191 Gif-sur-Yvette Cedex, France \\
$^{4}$Instituto de Astrof\'isica de Canarias, E-38205 La Laguna, Tenerife, Spain \\
$^{5}$Departamento de Astrofisica, Universidad de La Laguna, E-38205 La Laguna, Spain
}

\date{Accepted XXX. Received YYY; in original form ZZZ}

\pubyear{2017}

\begin{document}
\label{firstpage}
\pagerange{\pageref{firstpage}--\pageref{lastpage}}
\maketitle

\begin{abstract}
Recently, several efforts have been made to identify the seismic signatures of flares and magnetic 
activity in the Sun and Sun-like stars. In this work, we have analyzed the disk-integrated velocity 
and intensity observations of the Sun obtained from the GOLF and VIRGO/SPM instruments, respectively, on 
board the {\em SOHO} space mission covering several successive flare events, for the period 
from 11 February 2011 to 17 February 2011, of which 11 February 2011 remained a relatively quiet day and 
served as a ``null test'' for the investigation. Application of the spectral analysis to these 
disk-integrated Sun-as-a-star velocity and intensity 
signals indicates that there is enhanced power of the global modes of oscillations
in the Sun during the flares, as compared to the quiet day. The GOLF instrument obtains velocity 
observations using the Na~I~D~lines which are formed in the upper solar photosphere, while the 
intensity data used in our analysis are obtained by VIRGO/SPM instrument at 862~nm, 
which is formed within 
the solar photosphere. Despite the fact that the two instruments sample different layers of the 
solar atmosphere using two different parameters (velocity v/s intensity), we have found that both 
these observations show the signatures of flare-induced global waves in the Sun. These results 
could suffice in identifying the asteroseismic signatures of stellar flares and magnetic activity 
in the Sun-like stars.

\end{abstract}




\section{Introduction}
\label{sect:intro}

In solar flares, large amount of energy is suddenly released in the solar atmosphere 
within the time-scales of a few tens of minutes. Apart from these high budget thermal 
radiations ($\sim$ $10^{32}$~erg), 
flares also produce large amount of energetic particles that move at very high 
speeds (e.g., \citealp{Miller1997, Aschwanden2002, Emslie2005, Saint2005, Benz2008, Holman2011, Aschwanden2016}). 
With the advent of new and improved measurement techniques, several efforts have been made to understand 
the mechanisms involved in these explosive events occurring in the solar environment 
(e.g., \citealp{Hudson1995, Aschwanden2002, Lin2006, Benz2008, Holman2011, Caspi2015, Benz2017}). 
It is generally believed that the magnetic energy 
stored in the solar active regions gets released during the flares and that is 
demonstrated in the form of emission of high budget thermal energies, variety of high energy electromagnetic radiations 
and energetic particles, and coronal mass ejections (CMEs) 
(\citealp{Hundhausen1984, Walker1988, Webb1995, Srivastava2000, Aschwanden2002, Holman2011, Aschwanden2017}, and references therein). 
Large flares also produce energetic waves in the solar atmosphere and beneath the visible surface, such 
as Moreton waves \citep{Moreton1960}, and sunquakes \citep{Kosovichev1998}, respectively.
The changes in the solar magnetic fields 
accompanying the flares show their signatures in the form of the evolution of photospheric 
and chromospheric magnetic fields (e.g., \citealp{Wang1994, Hagyard1999, Lara2000, Sudol2005, 
Metcalf2005, Petrie2010, Wang2010, Wang2015, Kleint2017, Wang2017}).

Soon after the discovery of global oscillation modes in the Sun 
\citep{Leighton1962, Ulrich1970, Leibacher1971, Deubner1974}, 
it was speculated by \citet{Wolff1972} that solar flares 
could excite the free modes of oscillations of the entire Sun. The assumption 
behind this opinion of \citet{Wolff1972} was that the flares would cause a thermal expansion 
that would drive a mechanical impulse due to a compression front moving sub-sonically 
into the solar interior. Since then, several efforts have been made to study the correlation 
between the solar transient events, such as solar flares and the coronal mass ejections, and 
the strength of global modes of oscillations in the Sun using data from various sources 
\citep{Chaplin1995, Foglizzo1998, Foglizzoetal1998, Gavry1999, Ambastha2006, Richardson2012}. 
All these studies could 
not {\bf uniquely} asses the relation between the strength of global oscillation modes in the 
Sun and the transient events.

However, \citet{Karoff2008} found a high correlation between the energy in the 
high-frequency part of the acoustic spectrum of the Sun above the acoustic cut-off frequency 
(the pseudo-mode region, \citealp{Garcia1998, Garcia1999}) and the solar soft X-ray flux during 
the flares. They used the Sun-as-a-star disk-integrated intensity observations of the Sun 
obtained by Variability of Solar IRradiance and Gravity (VIRGO; \citealp{Frohlich1995, Frohlich1997}) 
instrument onboard {\em Solar and Heliospheric Observatory} ({\em SOHO}; \citealp{Domingo1995}) 
space mission and the disk-integrated 
soft X-ray emission (1--8~\AA~energy~band) from the Sun observed by the {\em Geostationary Operational 
Environmental Satellite} ({\em GOES}; \citealp{Garcia1994}) satellites. 
\citet{Kumar2010} presented the analysis of the full-disk velocity
observations of the Sun obtained from the Michelson and Doppler Imager (MDI; \citealp{Scherrer1995}) 
and the Sun-as-a-star disk-integrated velocity observations of the Sun obtained by the 
Global Oscillations at Low Frequencies (GOLF; \citealp{Gabriel1995, Gabriel1997}) instruments onboard 
{\em SOHO} for a few flare events. Their results indicated an enhancement of high-frequency 
global waves in the Sun during the flares. \citet{Kumar2011} further showed post-flare 
enhancements in the global modes of oscillations in the Sun for a different flare event.

Recently, using the observations from {\em Kepler} spacecraft \citep{Borucki2010} 
of 73 flares on 39 solar-like stars, \citet{Karoff2014} concluded that the photometric variability
of solar-like stars showed significant changes at the time of the flares. Thus, if the flares in 
solar-like stars have indeed caused seismic signatures, it provides a genuine case to further 
investigate a causal relation between the strength of flares and the changes in the global 
oscillation modes in the Sun.

In this work, we investigate the influence of flares on the global modes of oscillations 
in the Sun using the Sun-as-a-star disk-integrated velocity and intensity observations 
obtained by the GOLF and 
VIRGO/SPM instruments, respectively, onboard the {\em SOHO} space mission for the period from 
11 February 2011 to 17 February 2011, which covered several successive flare events.

In the following Sections, we first describe the observational data and the methods 
of our analysis, which is then followed by the results, and the conclusions with a discussion.

\section{The Observational Data}
\label{sect:obs}

The Sun remained highly active during the period from 12 February 2011 to 17 February 2011, since 
these days were 
populated with several sequential flare events. 11 February 2011 was relatively a quiet day with 
only a few short duration minor flares. This provided us the opportunity to use the observations obtained on 
11 February 2011 to serve as a ``null test'' for our investigation related to flare-induced global waves 
in the Sun for the 
period from 12 February 2011 to 17 February 2011. We have chosen a time window of 8 hours 
on each day from 12 February 2011 to 17 February 2011 for our analysis in such a way that it covers the 
maximum number of flares on the given day. On contrary to this, the time window of 8 hours chosen on 
11 February 2011 does not cover any flare events and also this is well separated from any minor flares that 
took place on this day. Fortunately, the time window of 8 hours considered in our analysis is a reasonable time 
to detect any flare-induced standing waves in the Sun, considering the sound crossing time inside the Sun to be 
around 2~hours ($t~=~2R\sun/<c_s>$; where~$<c_s>~\approx~200~kms^{-1}$,~average sound speed inside the Sun). 
In Table~1, we summarize the details of the period of observations considered in our analysis and 
the class of soft X-ray flares with their peak time which were observed with the {\em GOES-15} satellite in 
the 1.0--8.0~\AA~band. We present below the details of velocity and intensity observations from GOLF 
and VIRGO/SPM, respectively, that have been used in our analysis.

\begin{table*}
\centering
\caption{Details of the period of observations considered in GOLF and VIRGO analysis and the 
solar flares during these observing periods as obtained by {\em GOES-15} satellite in soft X-ray (1.0--8.0~\AA) 
for the period 11 February 2011 to 17 February 2011.}
\begin{tabular}{lcr} 
\hline
Date & Period of observation (UT) & Class of soft X-ray flares with peak time (UT)\\
\hline
11 & 09:00-17:00 & No flares\\
12 & 08:00-16:00 & B2.3(10:51), B1.0(11:54), B2.1(12:22), C2.6(15:06)\\
13 & 12:00-20:00 & C1.1(12:36), C4.7(13:56), M6.6(17:38)\\
14 & 12:00-20:00 & C1.7(12:00), C9.4(12:53), C7.0(14:27), B7.9(16:53), M2.2(17:26), C6.6(19:30)\\
15 & 00:00-08:00 & C2.7(00:38), X2.2(01:56), C4.8(04:32)\\
16 & 06:00-14:00 & C2.2(06:22), M1.1(07:44), C9.9(09:11), C3.2(10:32), C1.0(12:02)\\
17 & 08:00-16:00 & C1.2(08:13), C1.9(09:30), C2.6(10:28), C2.2(10:44), C2.4(12:36), C1.3(15:54)\\
\hline
\end{tabular}
\end{table*}

\subsection{GOLF Velocity Observations}
\label{sect:golf}
The GOLF instrument obtains Sun-as-a-star disk-integrated line-of-sight velocity observations 
using the Na~I~D~lines with a 10~s cadence. However, we have used the calibrated velocity data 
from GOLF \citep{Garcia2005} rebinned for every 60~s to match with the VIRGO/SPM 
observations. A high-pass filter has been applied to these data to reduce the signals from rotation 
and other slowly varying solar features with a cut-off established at three days. 
It is to be noted that we observe normal evolution of velocity oscillations in the 
Sun in these GOLF observations for the entire period - 11 February 2011 to 17 February 2011.

\subsection{VIRGO/SPM Intensity Observations}
\label{sect:virgo}
The VIRGO/SPM instrument has basically three Sun photometers which obtain Sun-as-a-star 
disk-integrated intensity observations, separately, in the 5~nm spectral irradiance bands 
centered at 402~nm~(blue), 500~nm~(green), and 862~nm~(red) with a 60 s cadence. 
We have used in our analysis 
the calibrated intensity data from VIRGO/SPM obtained at 862~nm because of its proximity with the 
GOLF observations in the Na~I~D~lines as regards to their formation height in the solar 
atmosphere \citep{Jimenez2005}. These calibrated time series are first subjected to orbital 
corrections due to the satellite motion and then 
a 1-day moving mean filter with double smoothing (i.e., a triangular filter) is applied to reduce 
the signals from solar rotation and slow 
background variations. It is to be noted that time series used
in this work belong to a longer time series from which the interested days have been extracted to avoid the
border effects of the aforementioned filter. Finally, a two-point backward difference filter 
is applied to 
the time series to reduce the low-frequency residuals. This improves the visibility of the 
high-frequency peaks in the spectrum. Similar to the GOLF observations, we observe normal 
evolution of intensity oscillations in the Sun in these VIRGO/SPM observations for the entire 
period - 11 February 2011 to 17 February 2011.

\section{Analysis and Results}
\label{sect:results}

\begin{figure*}
\centering
\includegraphics[width=0.4444\textwidth]{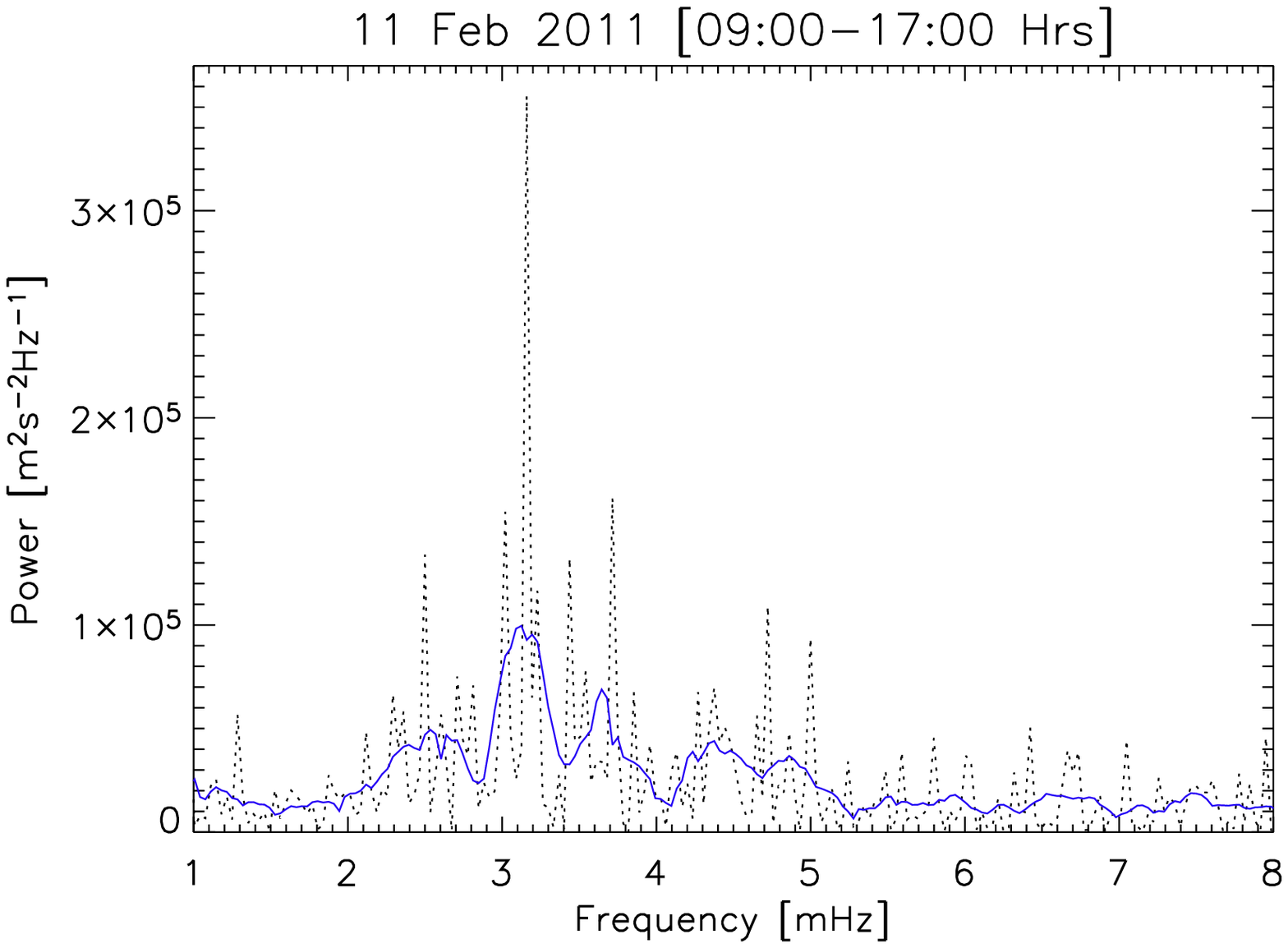} \hspace*{0.34 cm} 
\includegraphics[width=0.4444\textwidth]{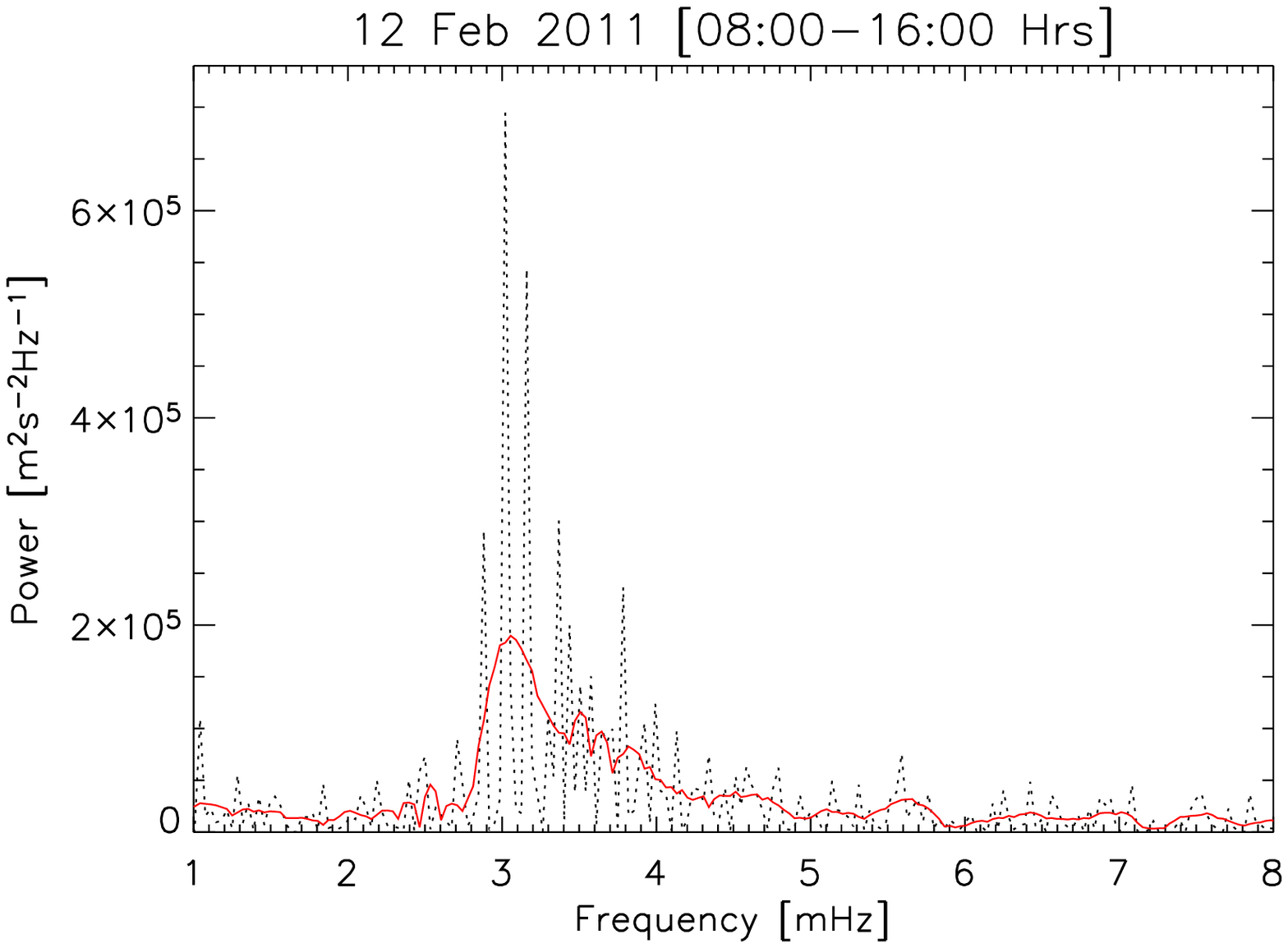} \\
\vspace*{0.07 cm}
\includegraphics[width=0.4444\textwidth]{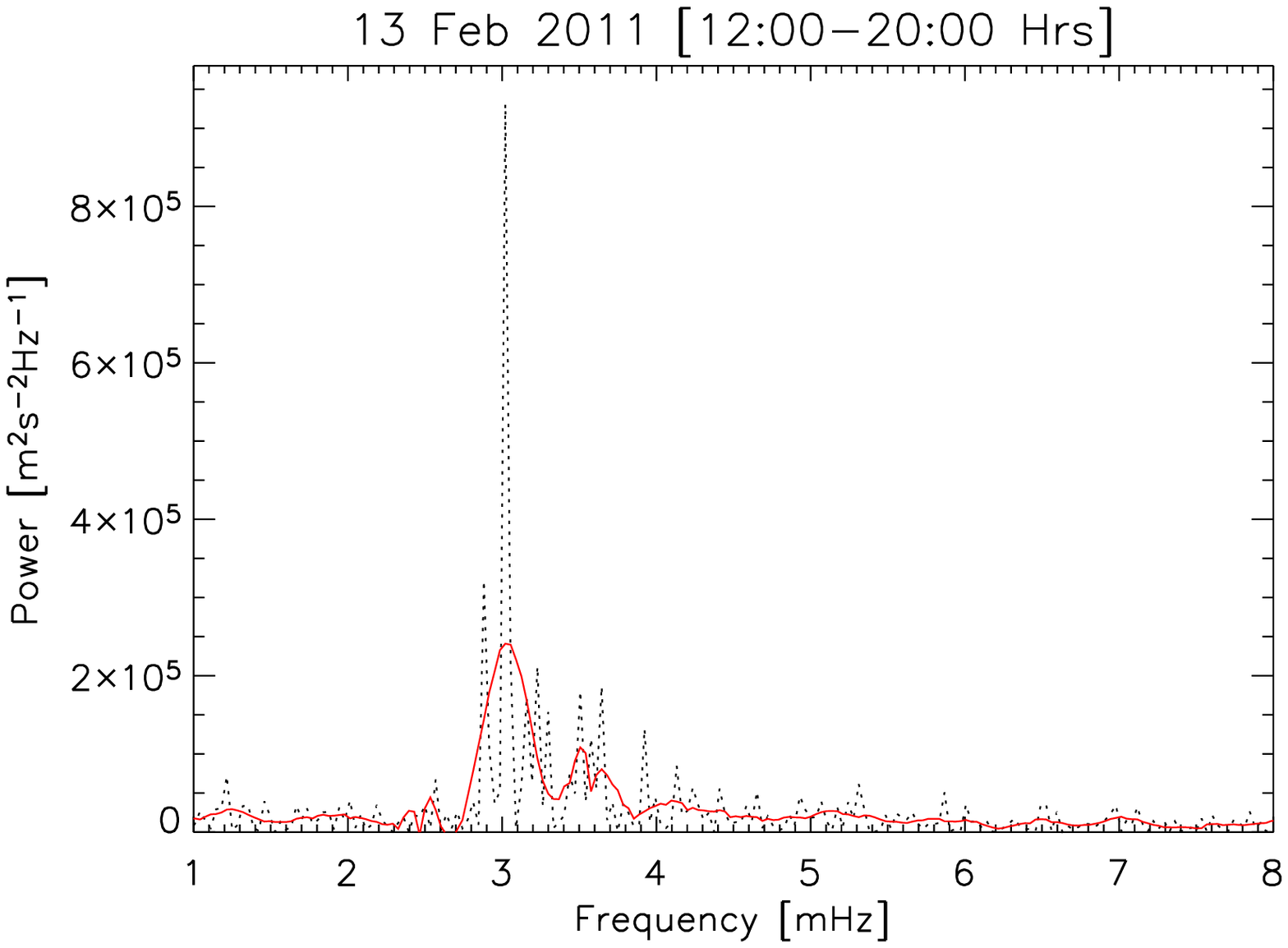} \hspace*{0.34 cm} 
\includegraphics[width=0.4444\textwidth]{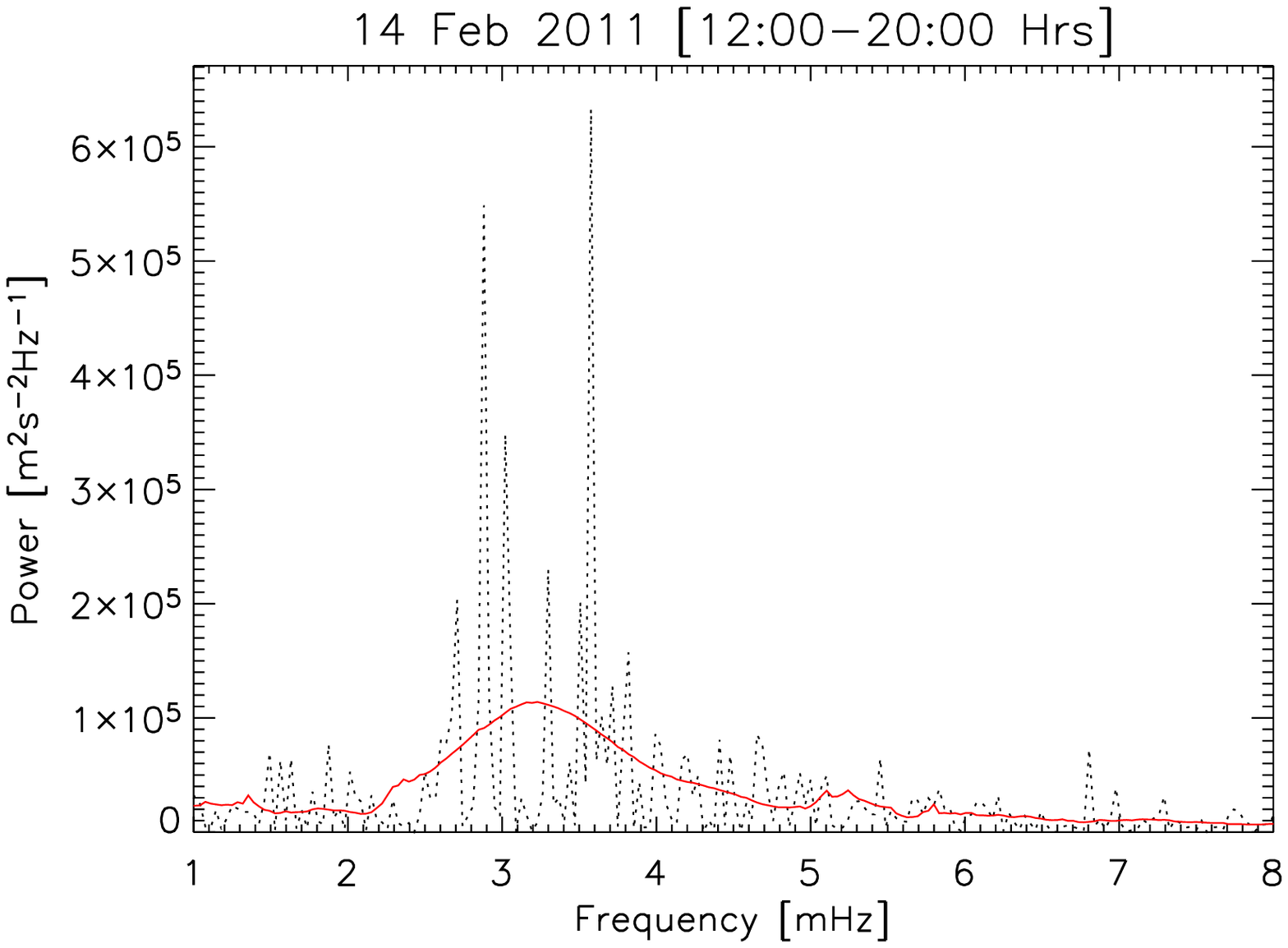} \\
\vspace*{0.07 cm}
\includegraphics[width=0.4444\textwidth]{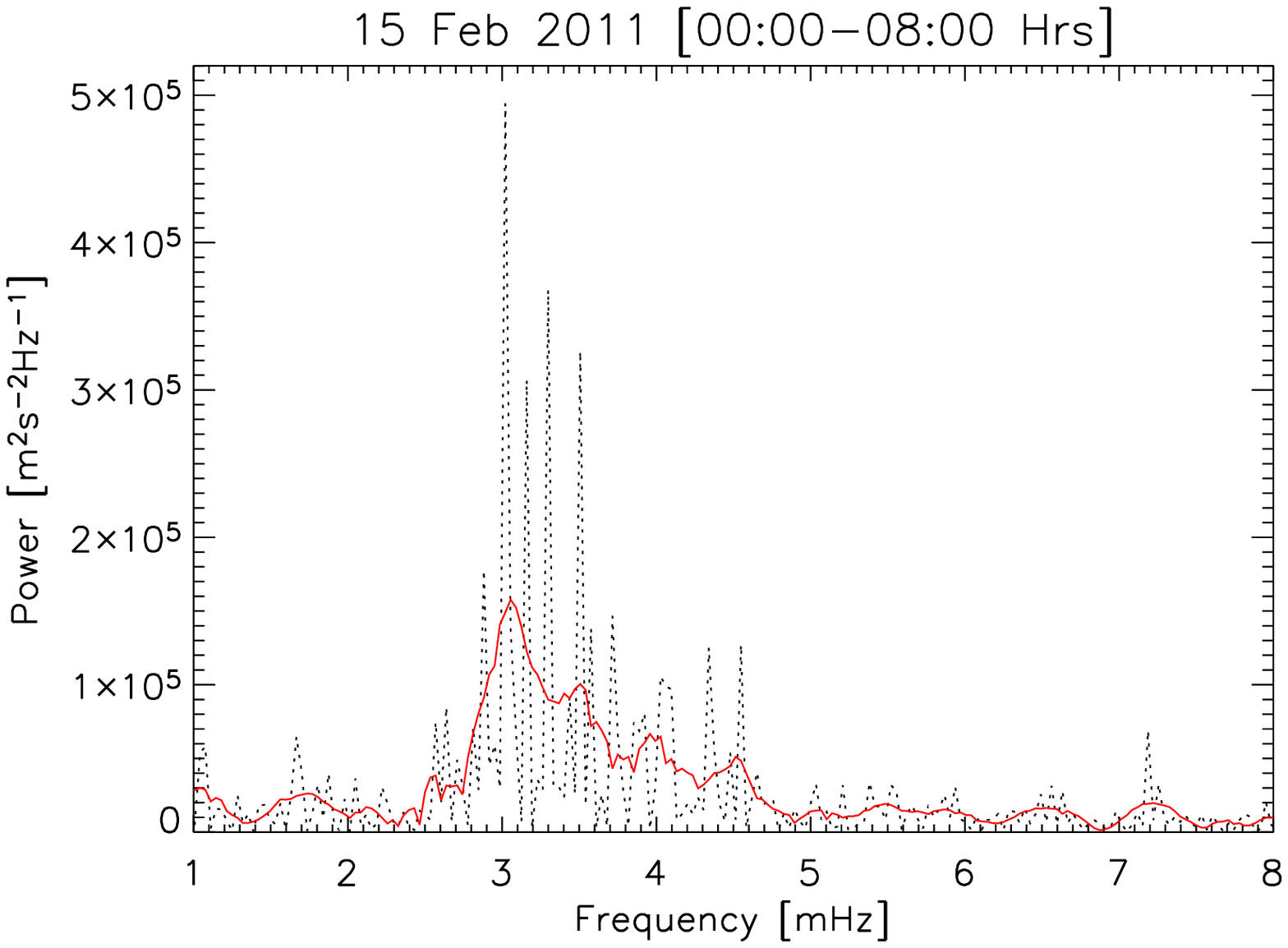} \hspace*{0.34 cm} 
\includegraphics[width=0.4444\textwidth]{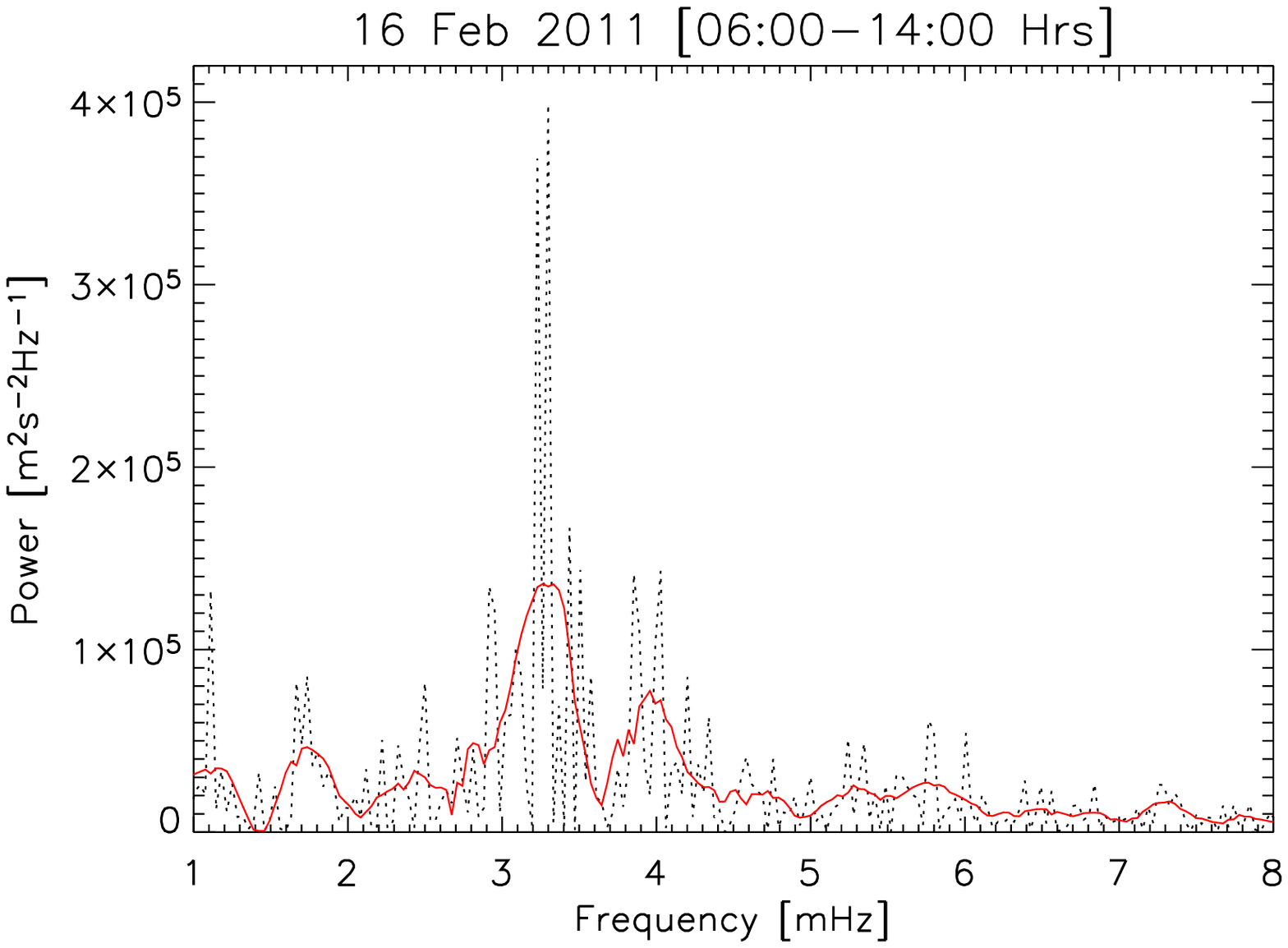} \\
\vspace*{0.07 cm}
\includegraphics[width=0.4444\textwidth]{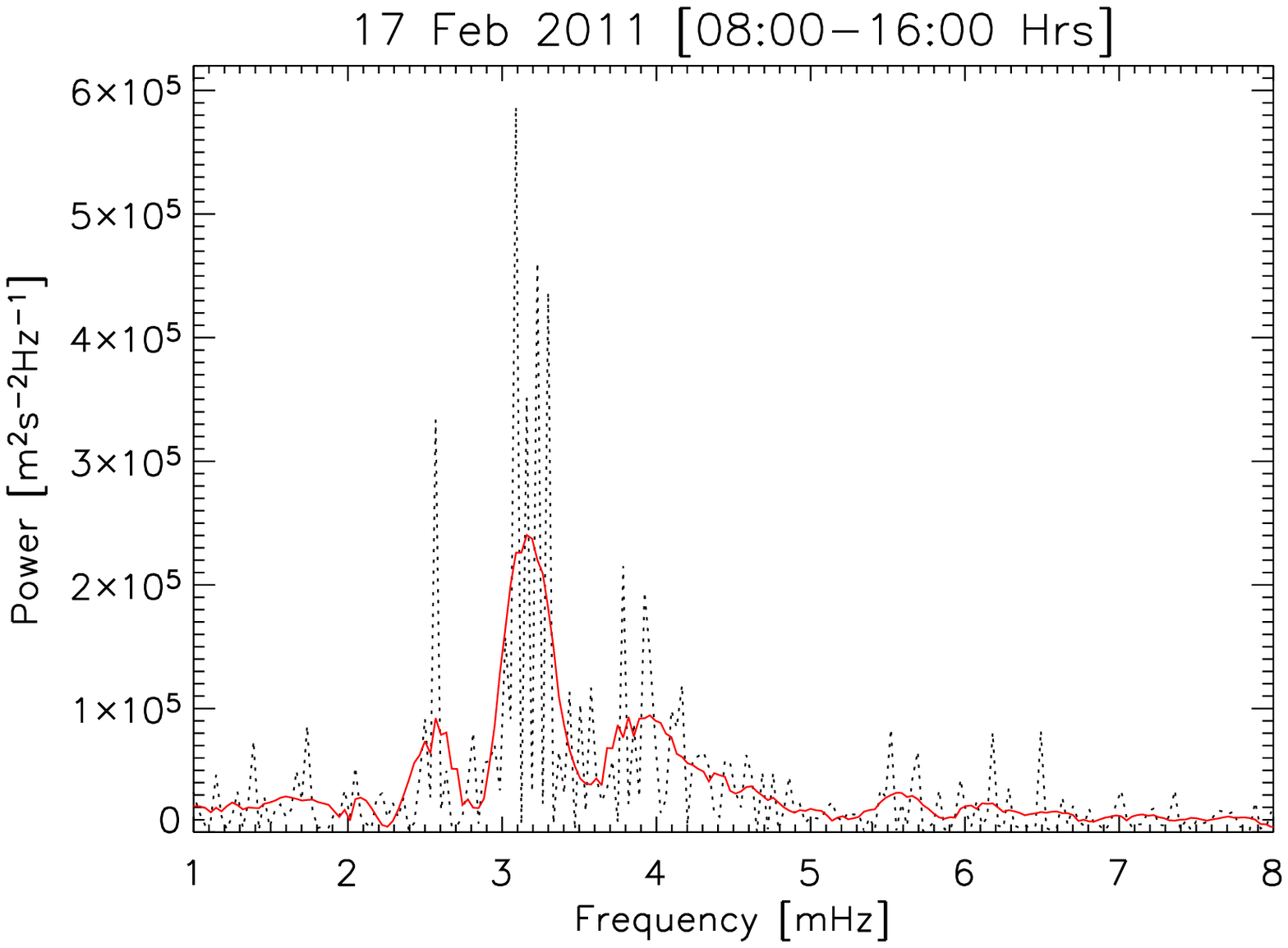}
\caption{Dashed black lines in all the panels are the 
Fourier Power Spectra of the Sun-as-a-star velocity observations from GOLF 
instrument using the Na~I~D~lines for the period 11 February 2011 to 17 February 2011. 
The plots shown in solid blue/red lines in all 
the panels represent a smoothing fit (S-G Fit) applied to the original power spectrum 
to estimate the power envelopes.}
\end{figure*}

\begin{figure*}
\centering
\includegraphics[width=0.4444\textwidth]{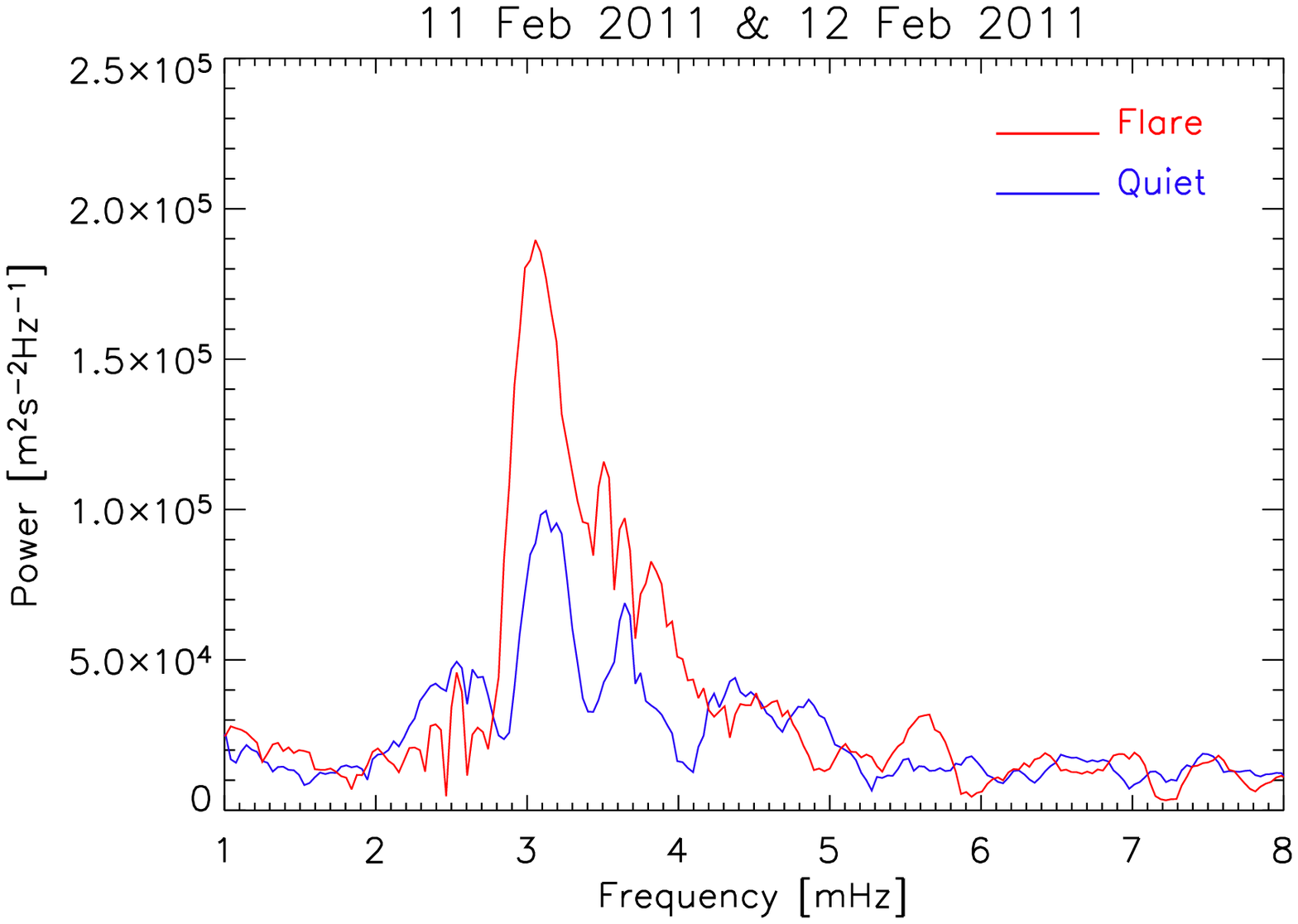} \hspace*{0.34 cm}
\includegraphics[width=0.4444\textwidth]{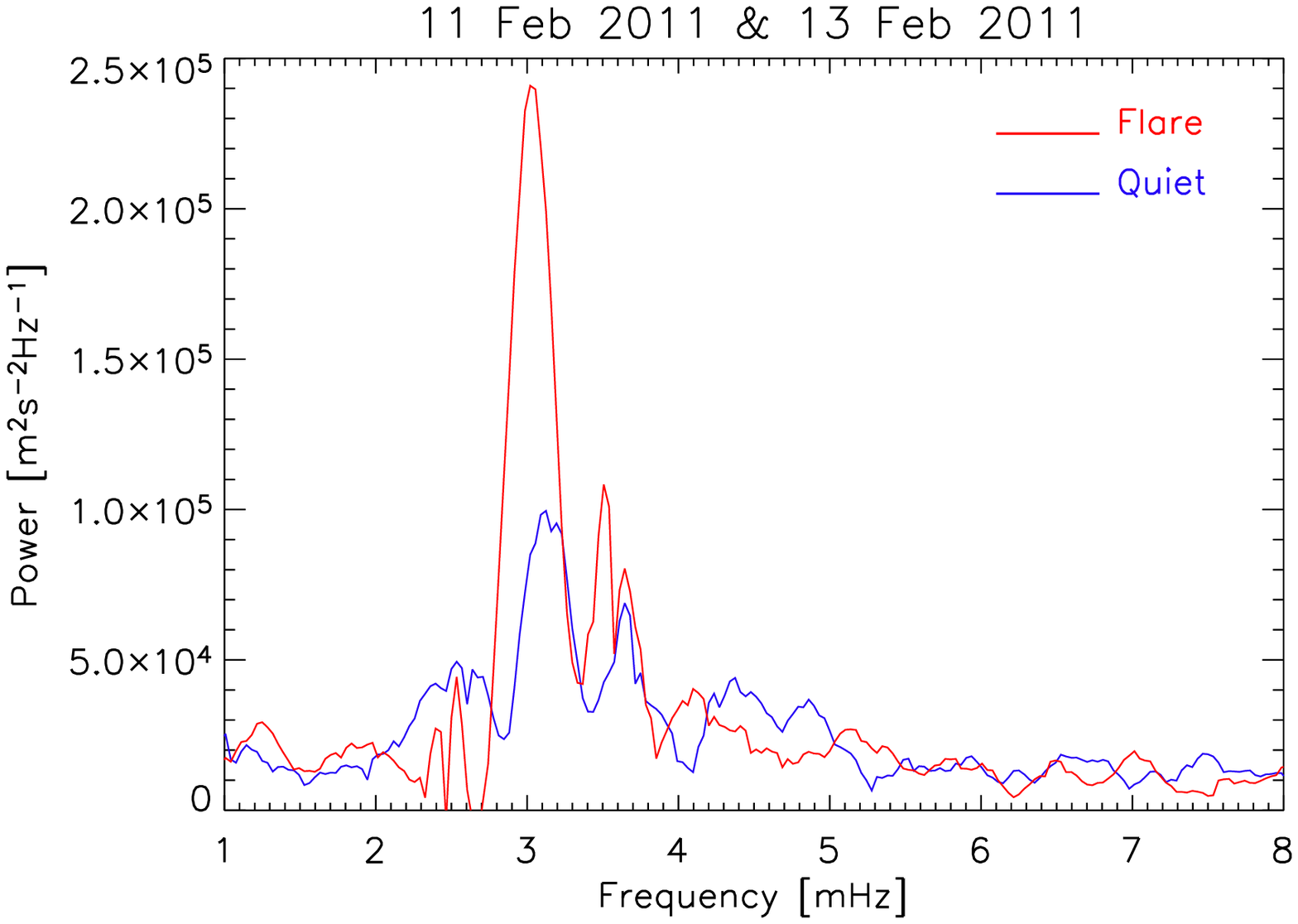}  \\
\vspace*{0.07 cm}
\includegraphics[width=0.4444\textwidth]{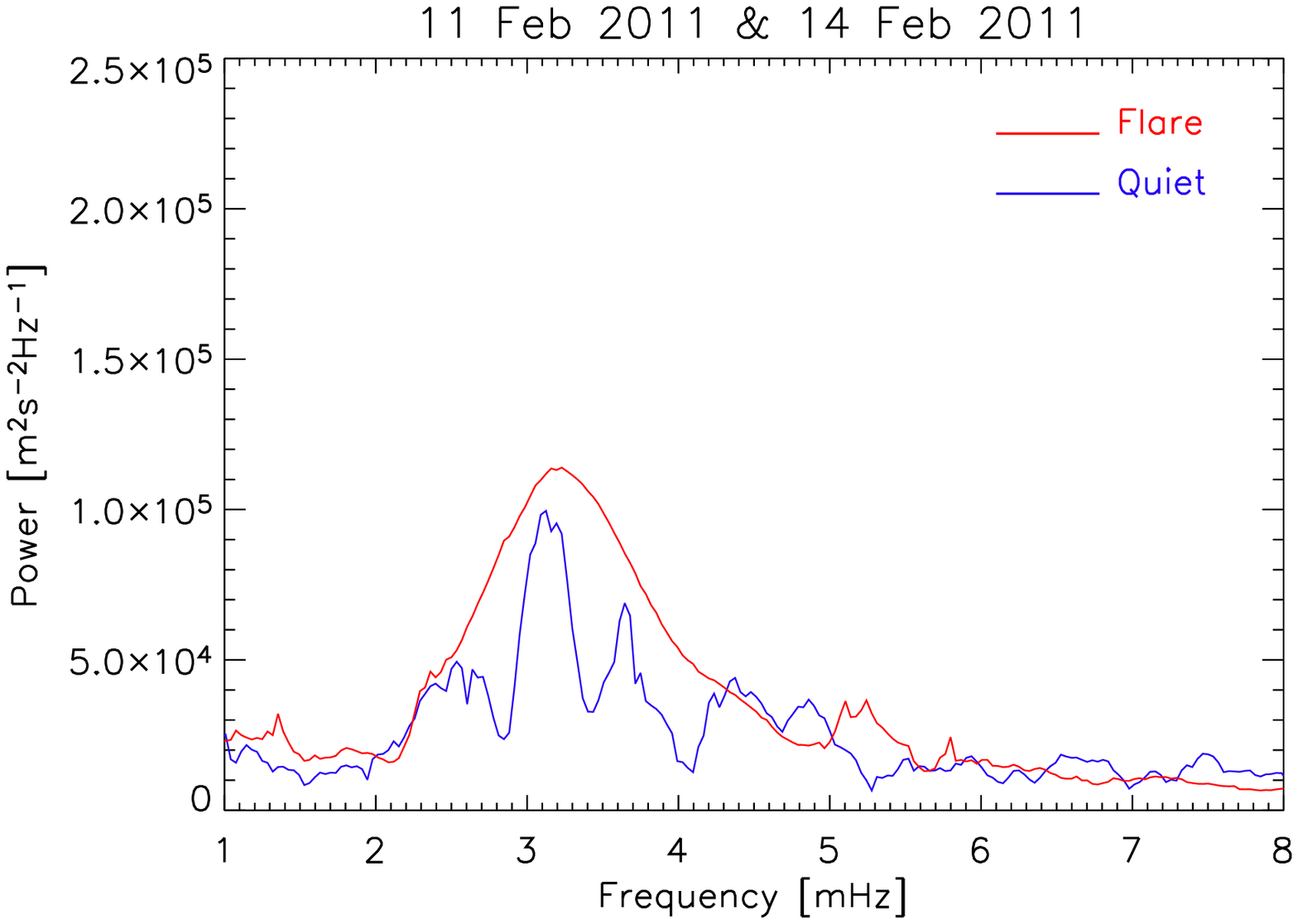} \hspace*{0.34 cm}
\includegraphics[width=0.4444\textwidth]{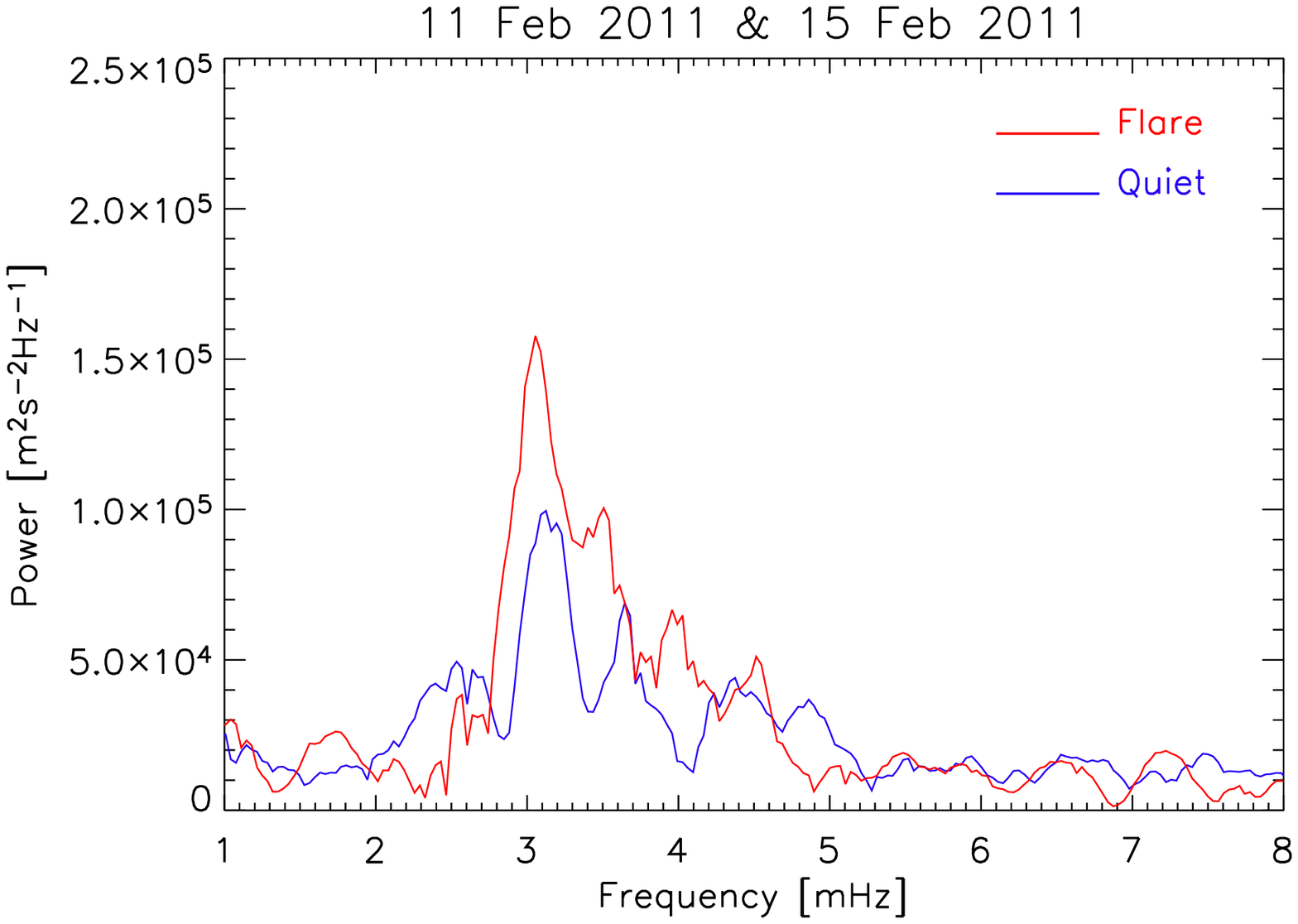}  \\
\vspace*{0.07 cm}
\includegraphics[width=0.4444\textwidth]{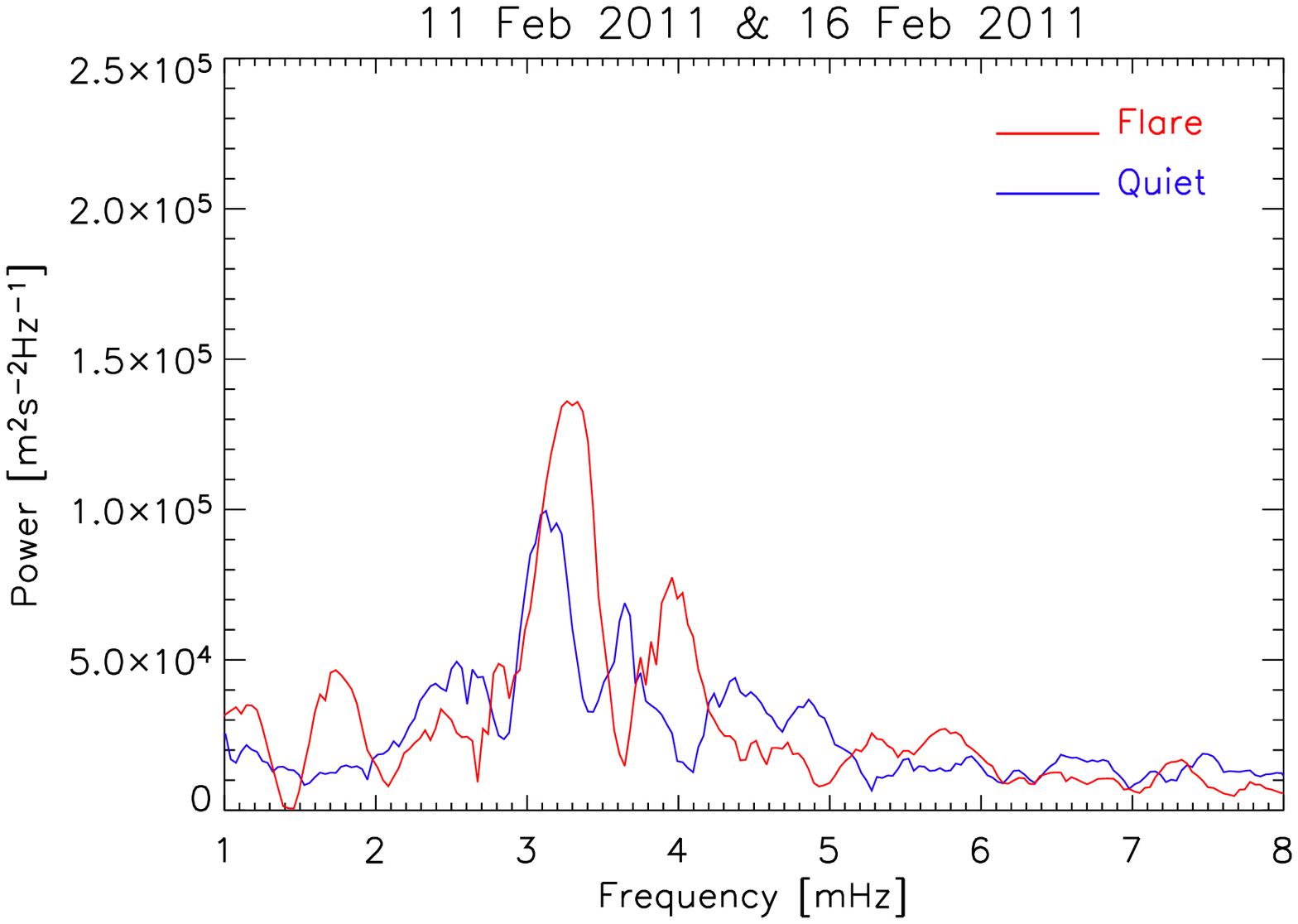} \hspace*{0.34 cm}
\includegraphics[width=0.4444\textwidth]{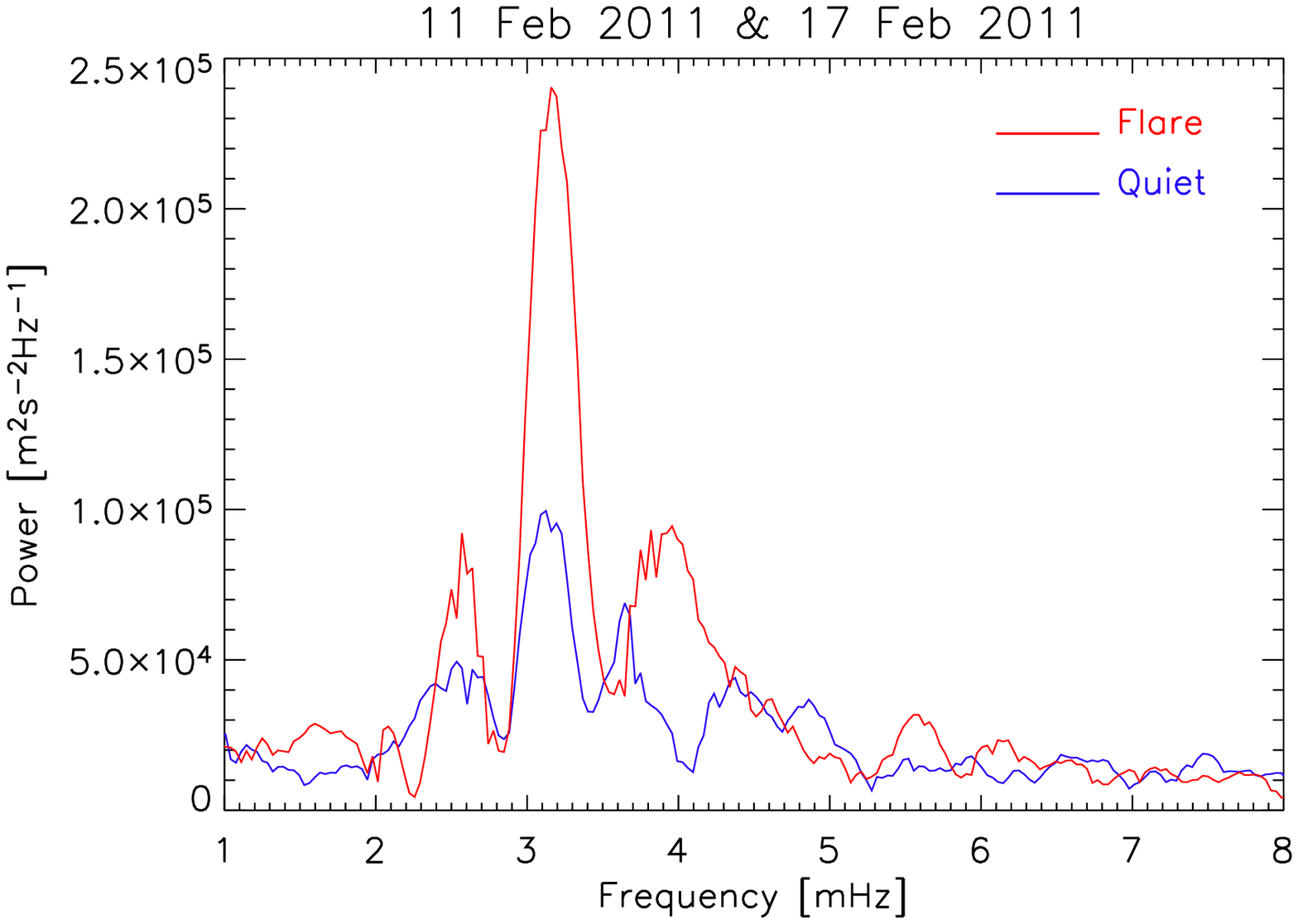}
\caption{Comparison of smoothed (S-G Fit) Fourier Power Spectra obtained from GOLF velocity 
observations (c.f., Figure~1) for the quiet day (11 February 2011 as shown in blue lines) and the days 
populated with several successive flare events (12-17 February, 2011 as shown in red lines). It is observed 
that the power of oscillatory modes are enhanced in different frequency regimes on the days having 
flares as compared to the quiet day.}
\end{figure*}

\begin{figure*}
\centering
\includegraphics[width=0.61\textwidth]{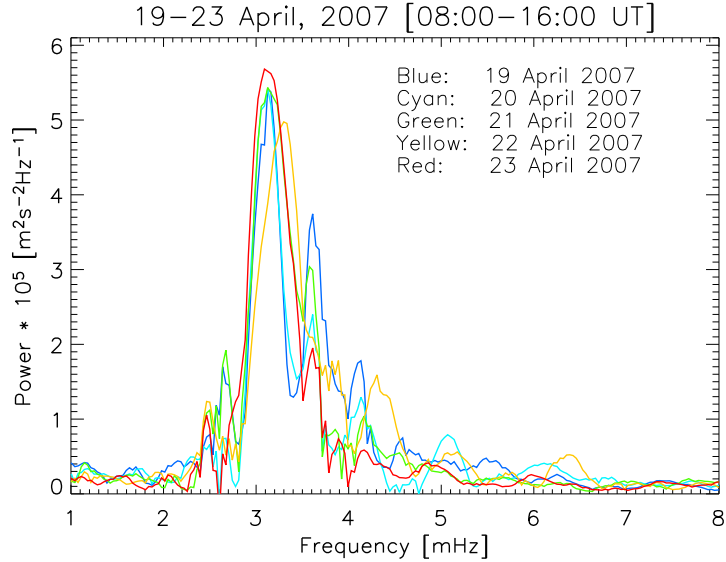}
\caption{Comparison of smoothed (S-G Fit) Fourier Power Spectra obtained from GOLF velocity 
observations for the 
period 19 April 2007 to 23 April 2007 during the extended solar magnetic-activity minimum 
with no reported flares. It is observed 
that these power spectra, over a period of five days, do not show the extent of variations as seen 
in the case of flare days with respect to the quiet day from GOLF observations (c.f., Figure~2).}
\end{figure*}

\begin{figure*}
\centering
\includegraphics[width=0.4444\textwidth]{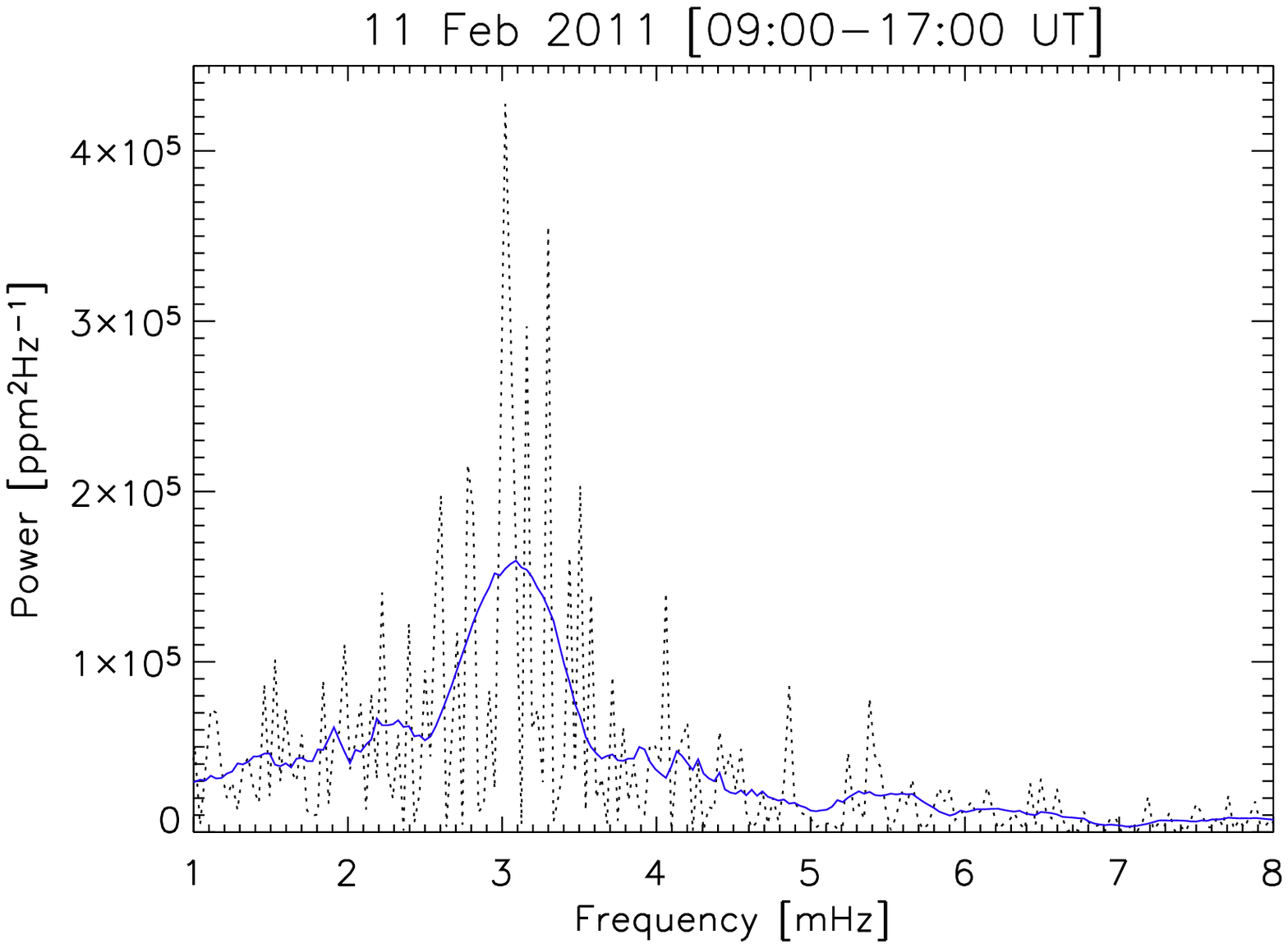} \hspace*{0.34 cm}
\includegraphics[width=0.4444\textwidth]{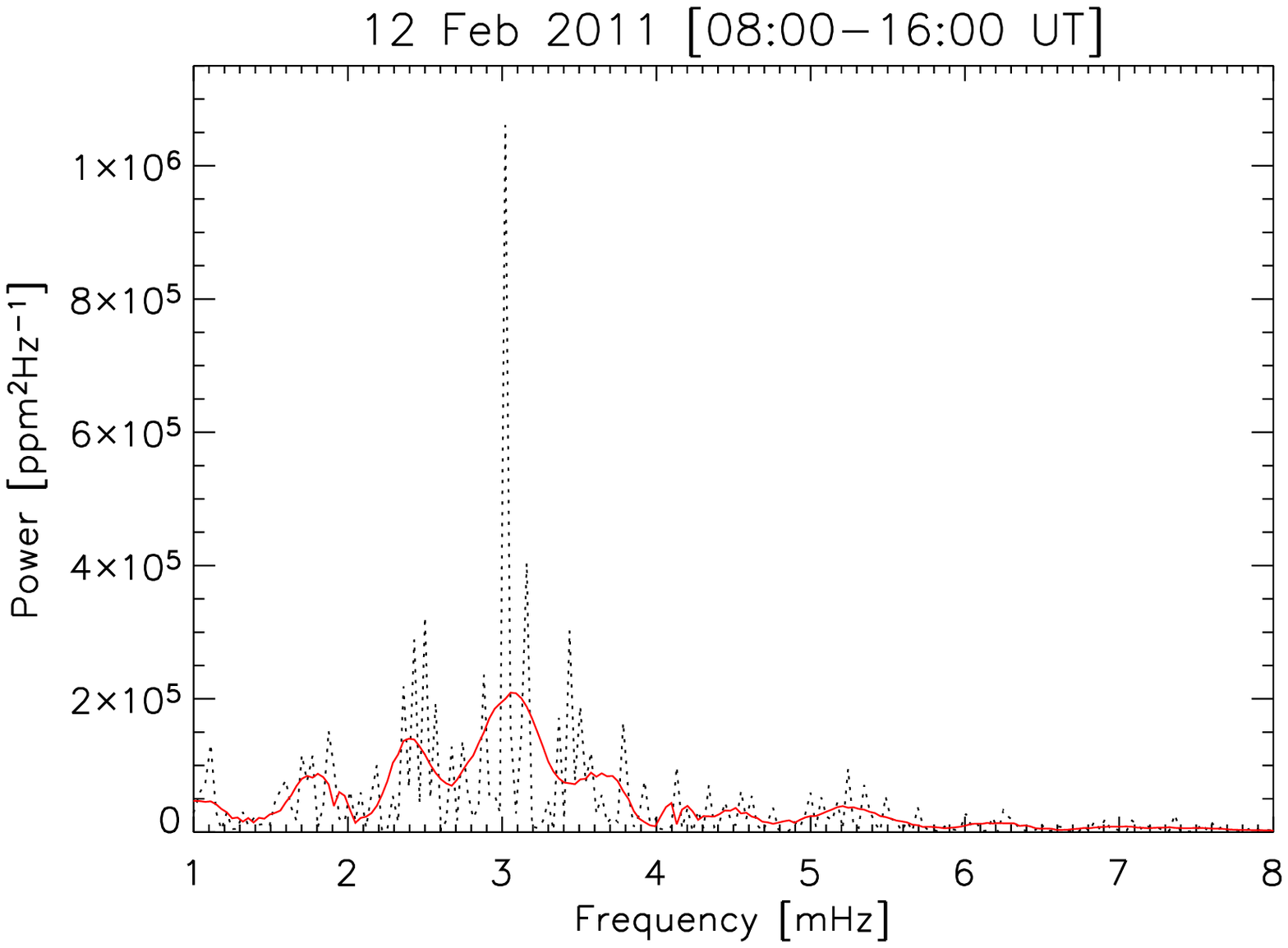}  \\
\vspace*{0.07 cm}
\includegraphics[width=0.4444\textwidth]{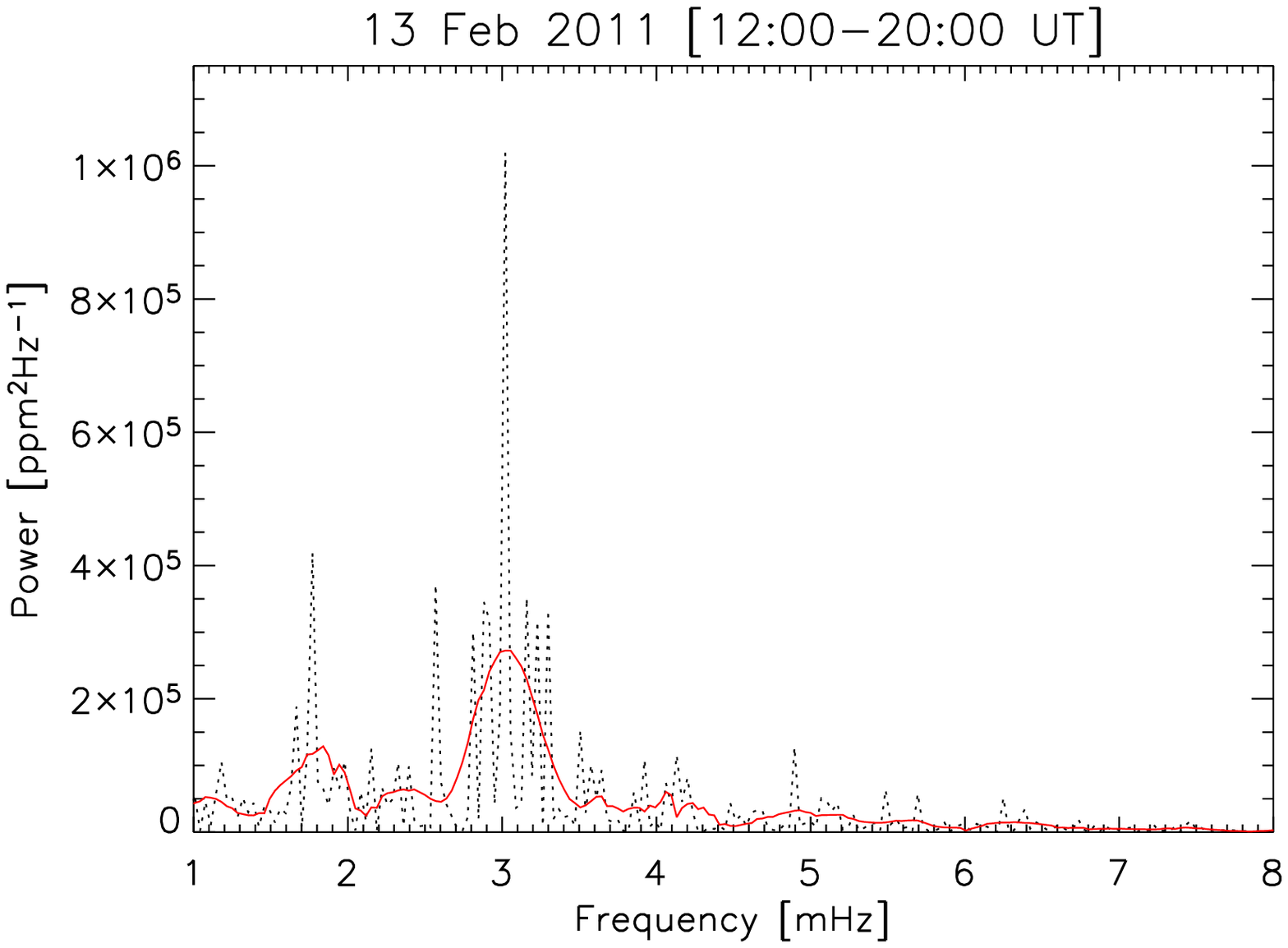} \hspace*{0.34 cm}
\includegraphics[width=0.4444\textwidth]{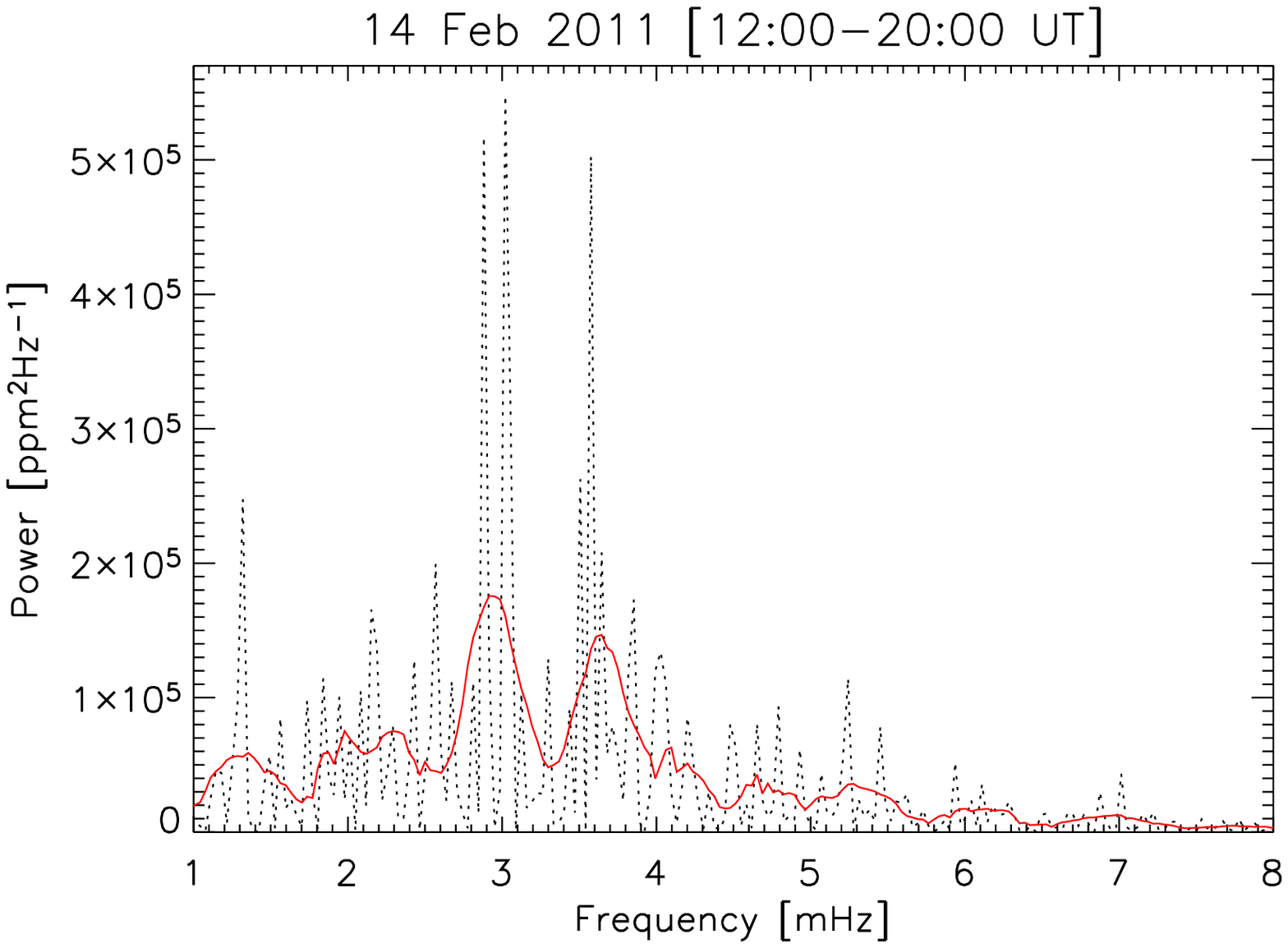}  \\
\vspace*{0.07 cm}
\includegraphics[width=0.4444\textwidth]{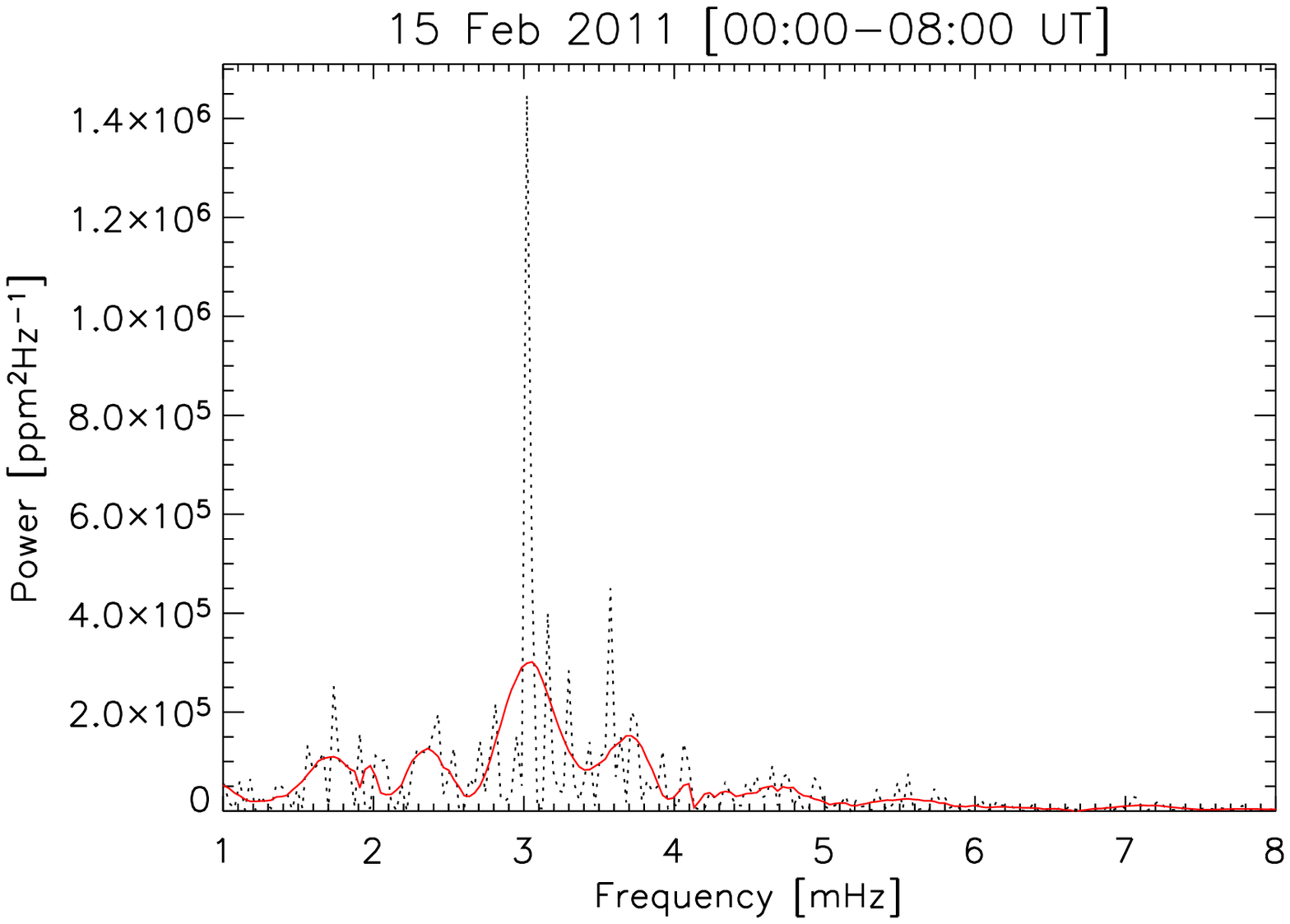} \hspace*{0.34 cm}
\includegraphics[width=0.4444\textwidth]{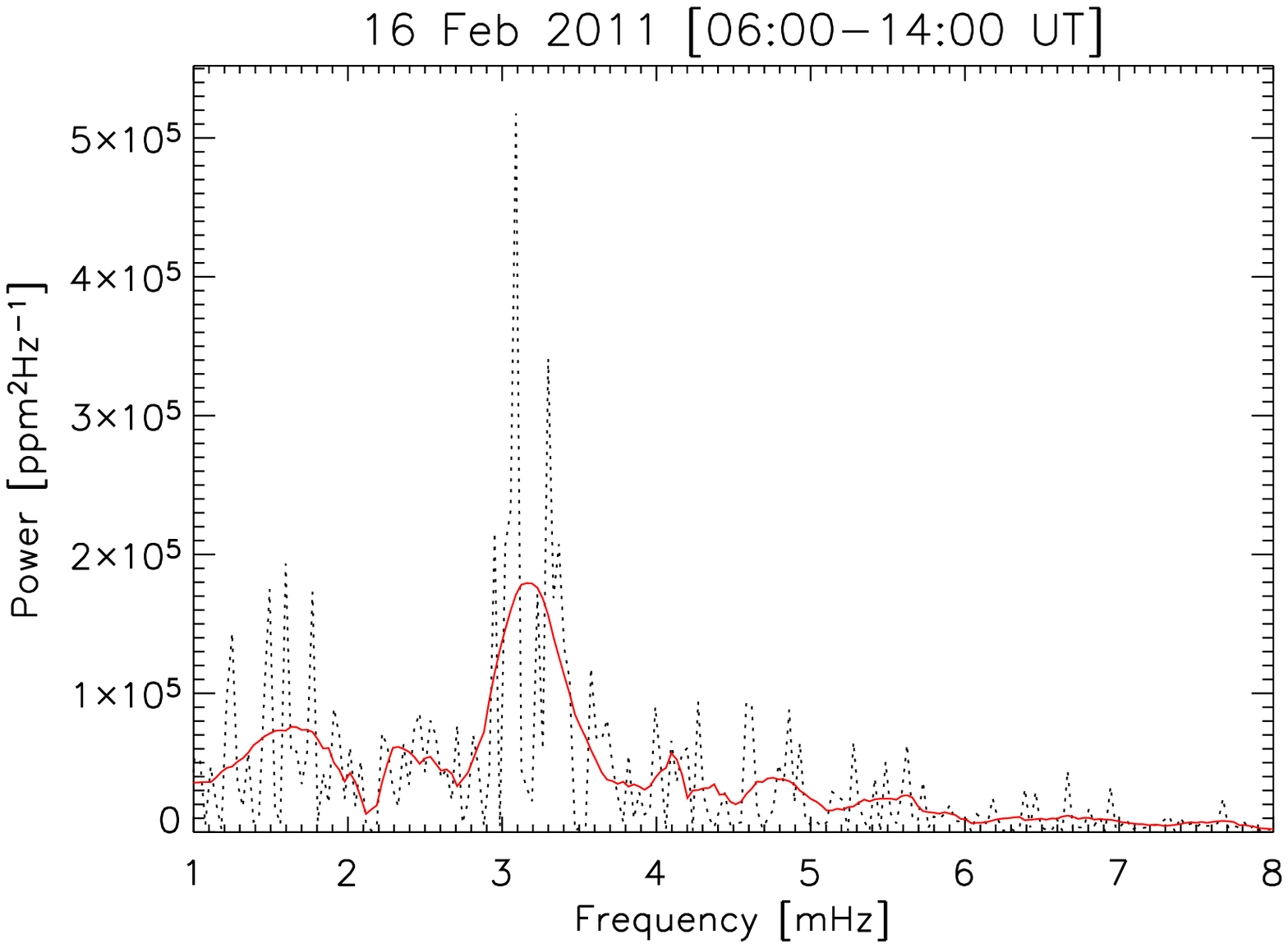}  \\
\vspace*{0.07 cm}
\includegraphics[width=0.4444\textwidth]{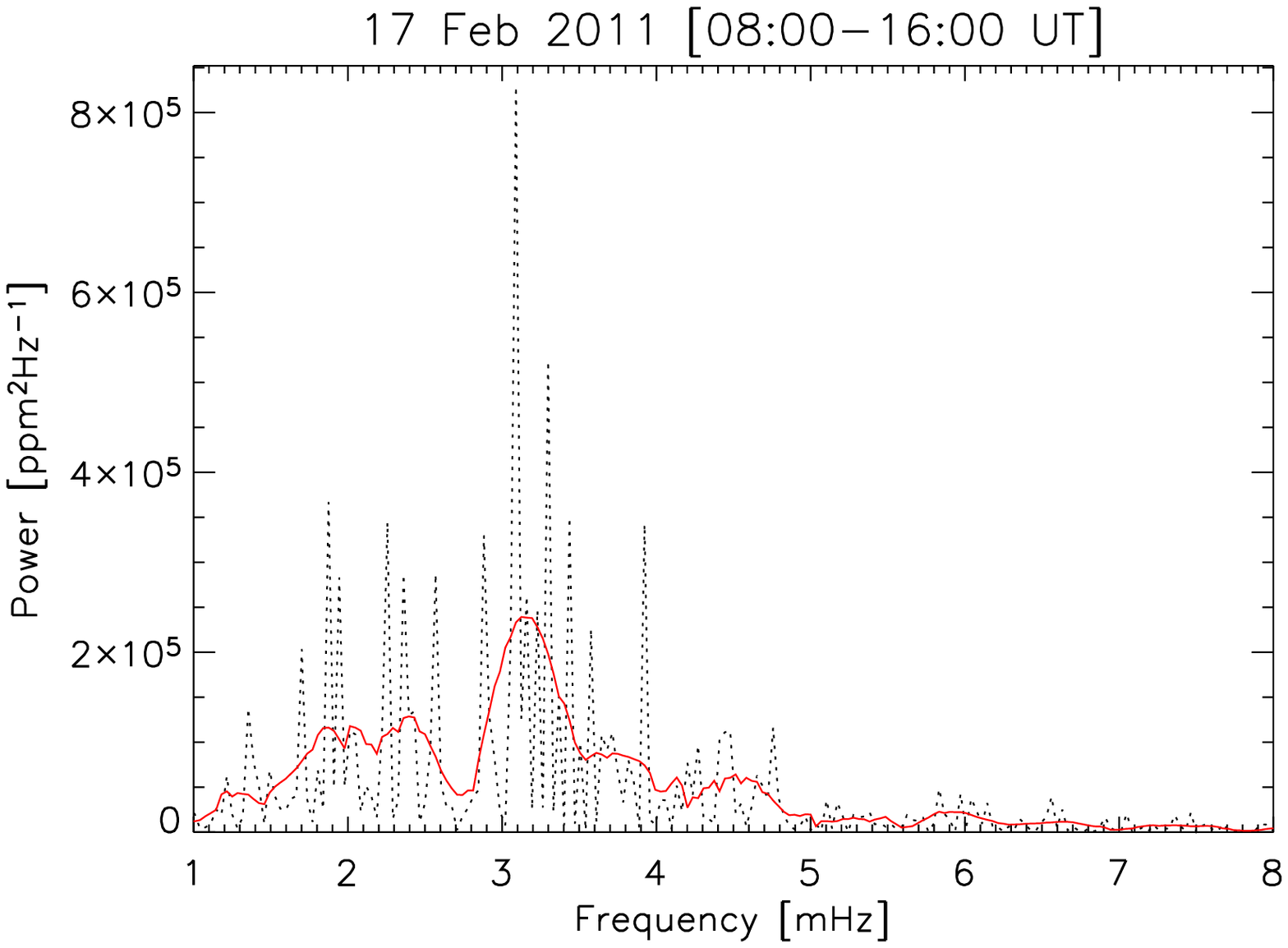}
\caption{Dashed black lines in all the panels are the 
Fourier Power Spectra of the Sun-as-a-star intensity observations from VIRGO/SPM instrument at 862~nm 
for the period 11 February 2011 to 17 February 2011. The plots shown in solid blue/red lines in all 
the panels represent a smoothing fit (S-G Fit) applied to the original power spectrum 
to estimate the power envelopes.}
\end{figure*}

\begin{figure*}
\centering
\includegraphics[width=0.4444\textwidth]{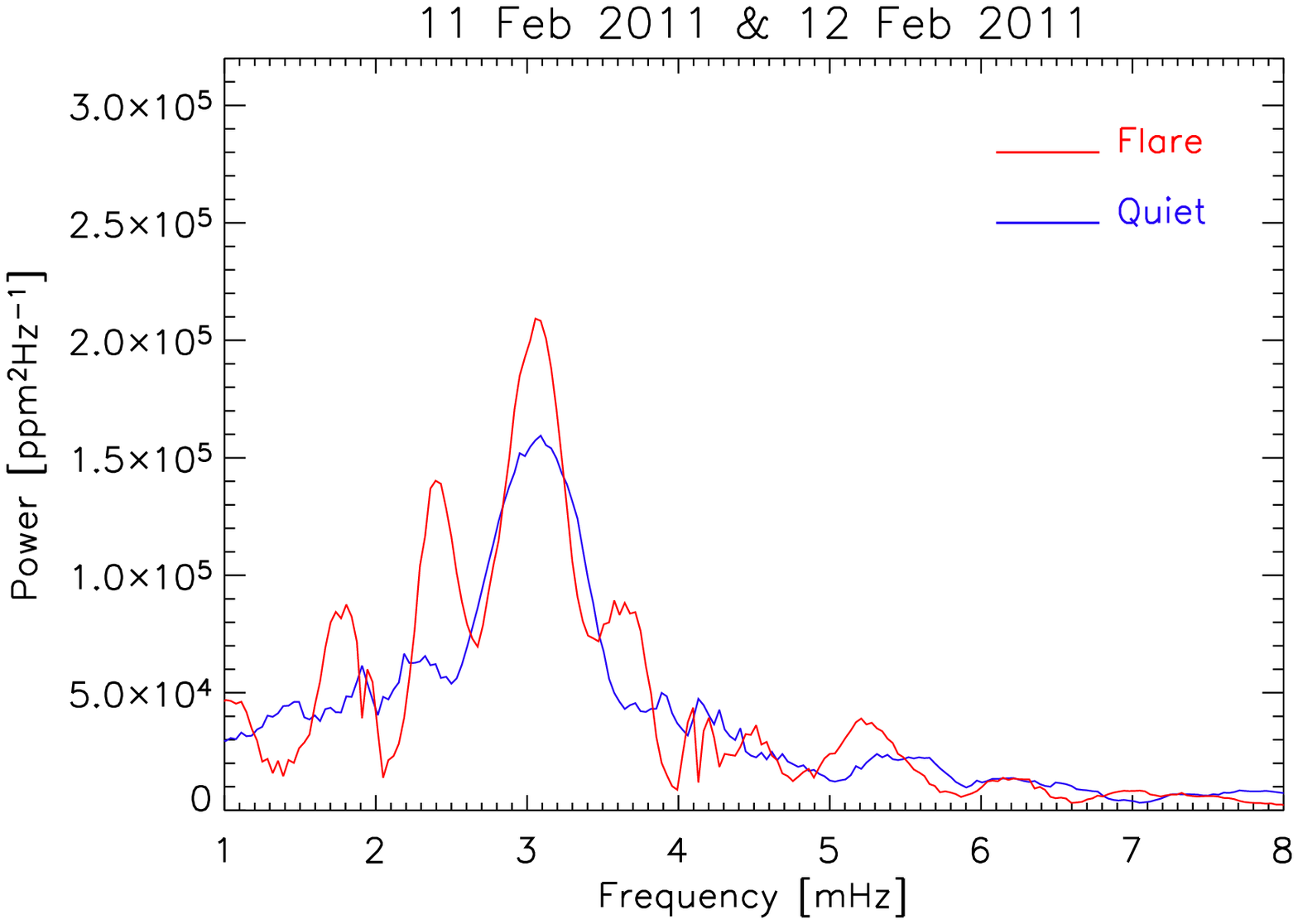} \hspace*{0.34 cm}
\includegraphics[width=0.4444\textwidth]{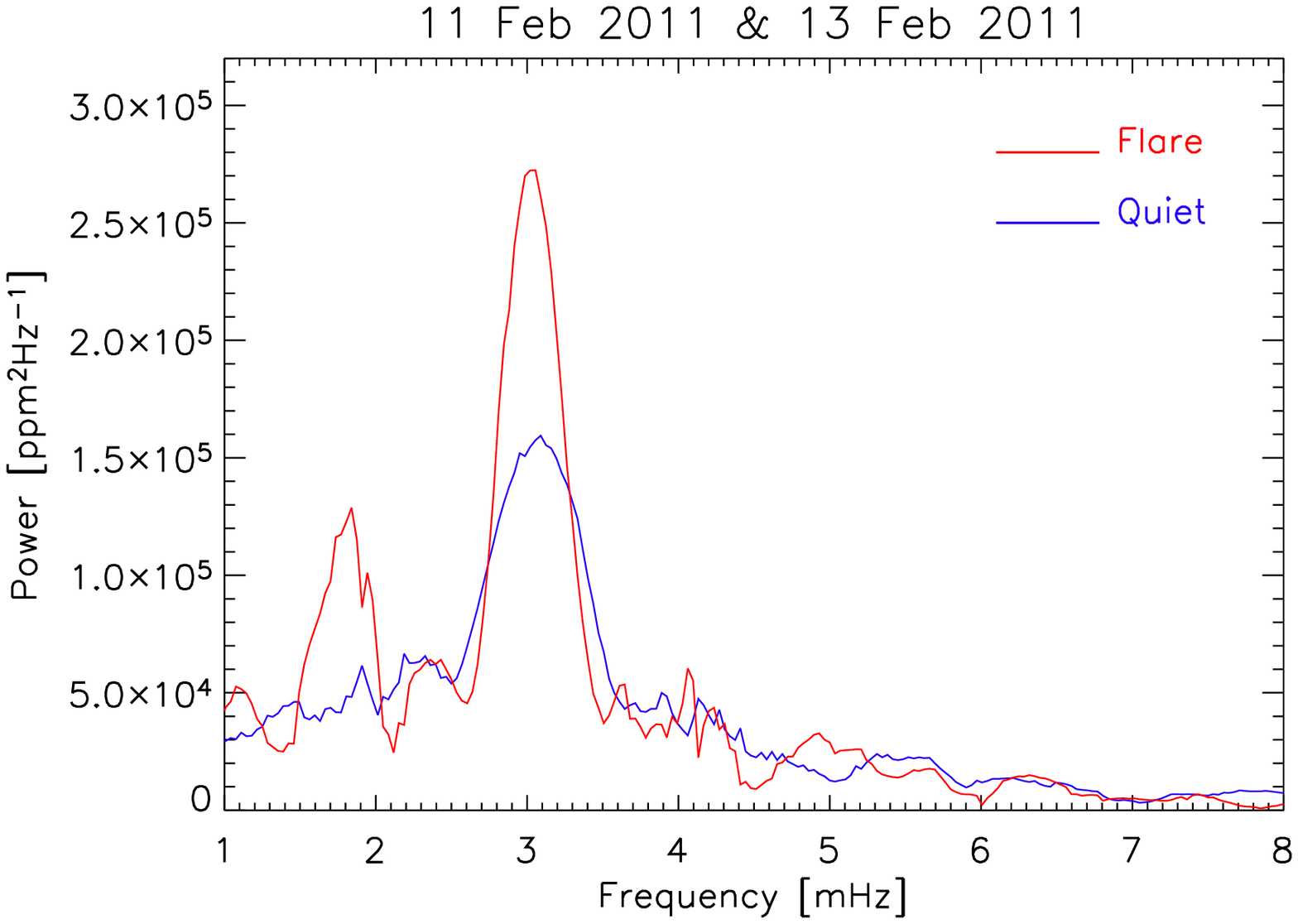}  \\
\vspace*{0.07 cm}
\includegraphics[width=0.4444\textwidth]{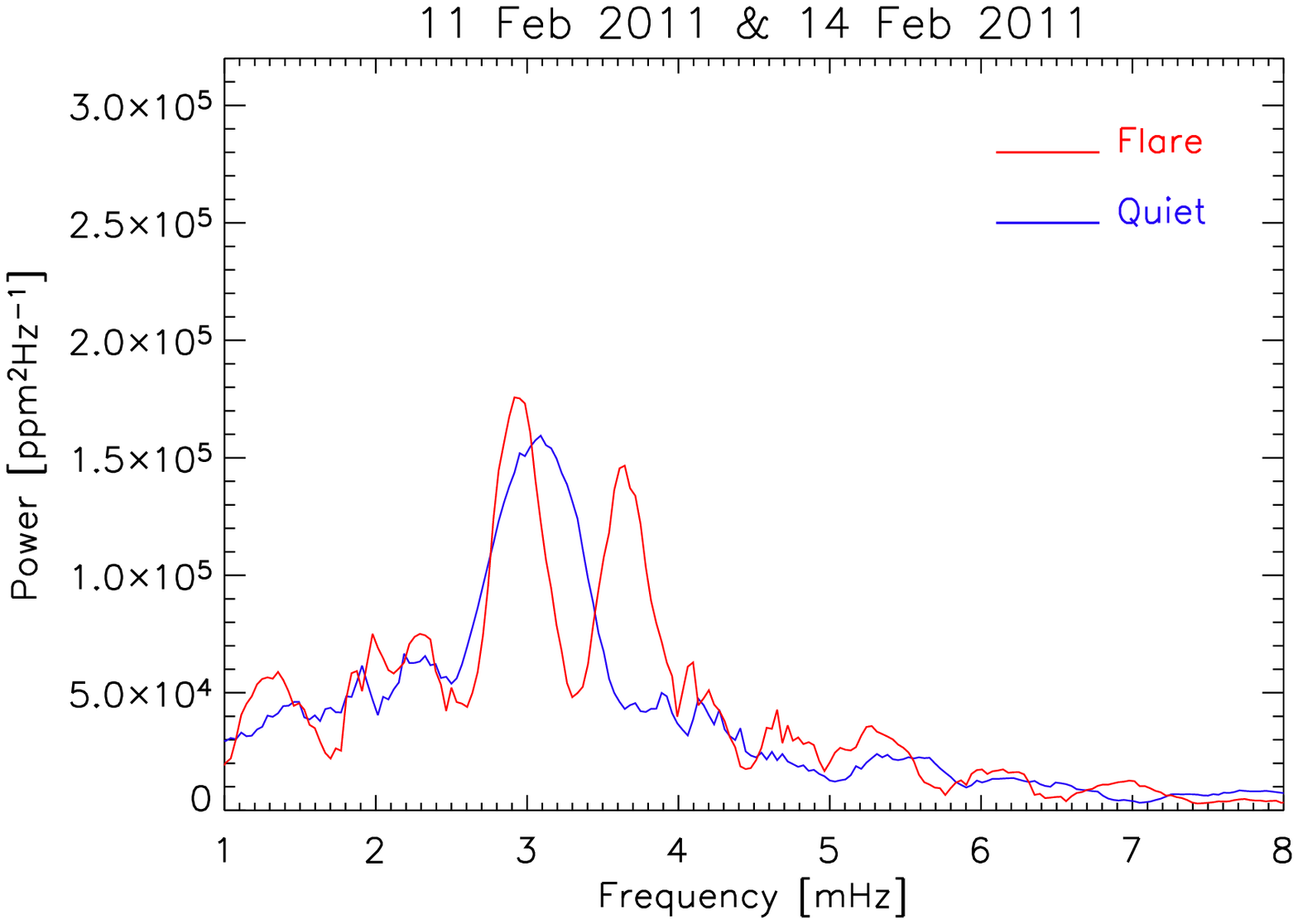} \hspace*{0.34 cm}
\includegraphics[width=0.4444\textwidth]{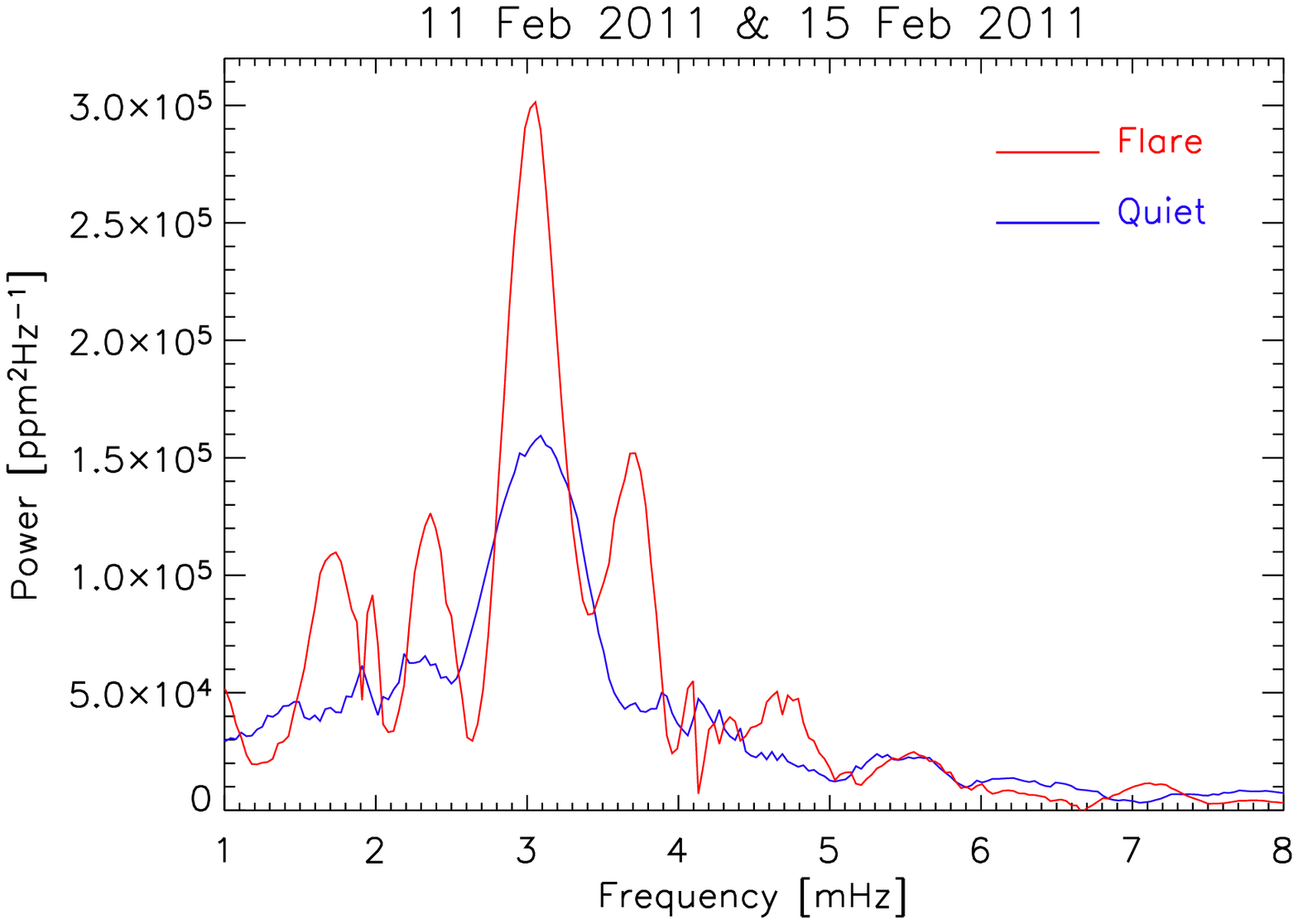}  \\
\vspace*{0.07 cm}
\includegraphics[width=0.4444\textwidth]{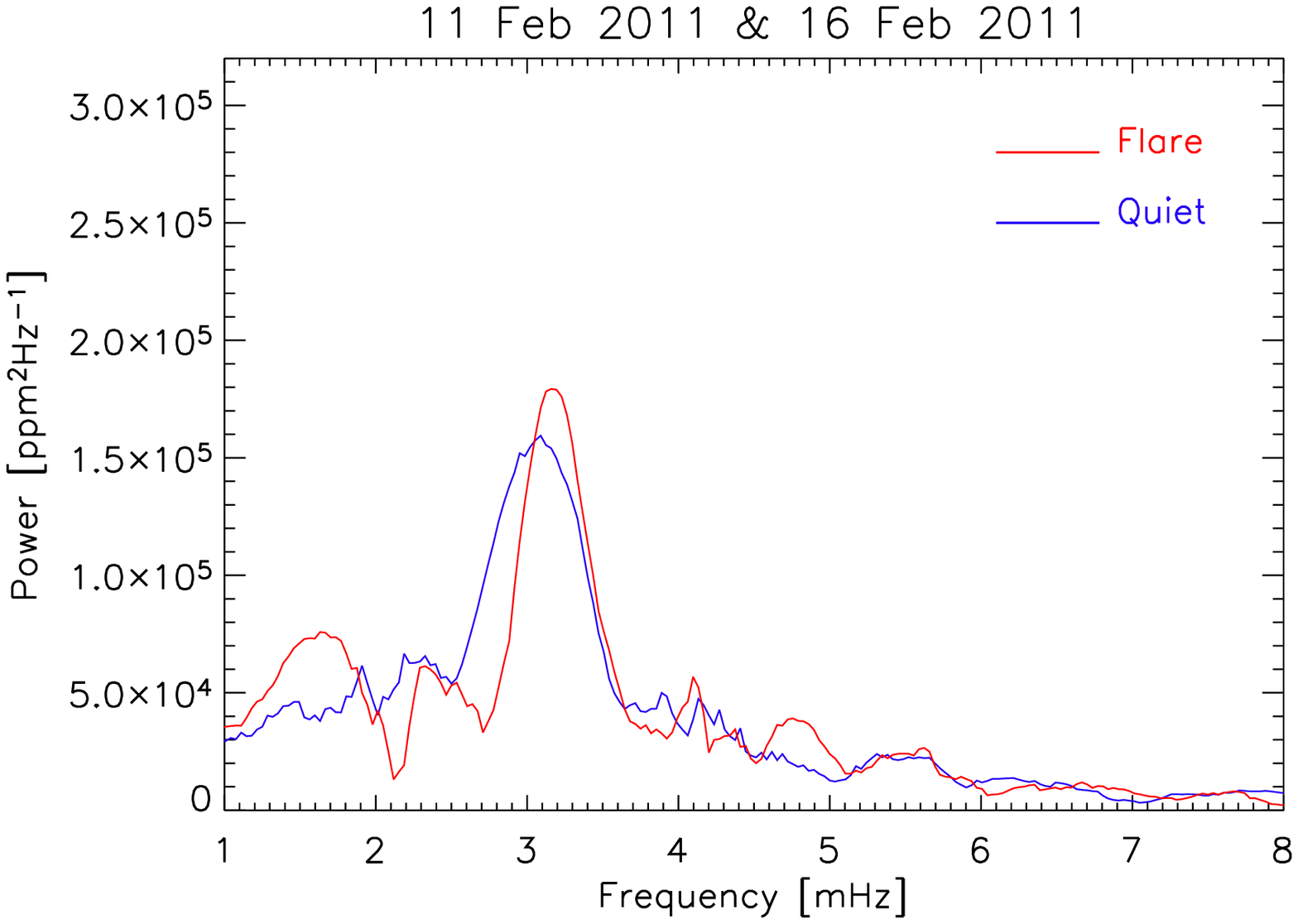} \hspace*{0.34 cm}
\includegraphics[width=0.4444\textwidth]{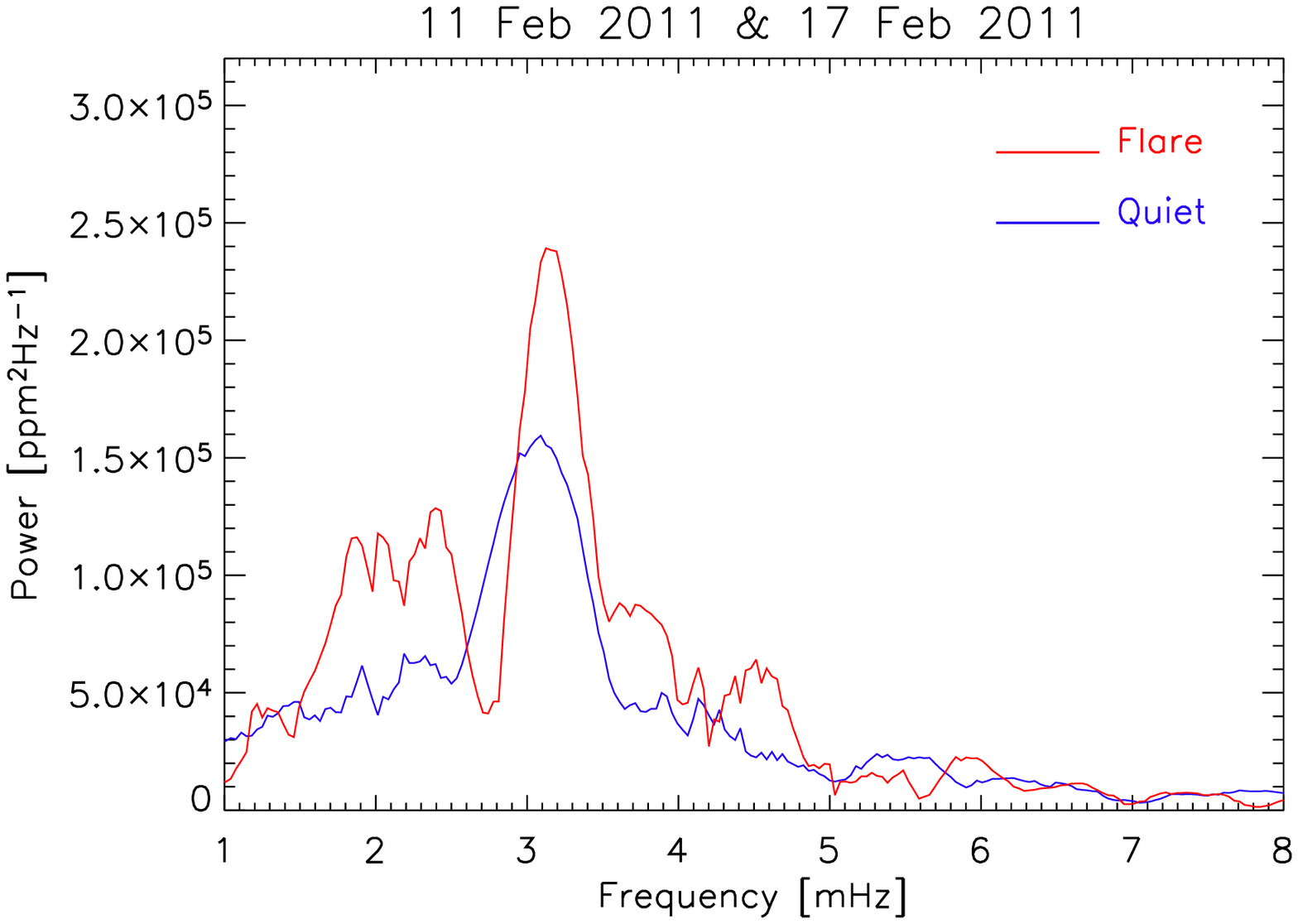}
\caption{Comparison of smoothed (S-G Fit) Fourier Power Spectra obtained from VIRGO intensity 
observations (c.f., Figure~4) for the quiet day (11 February 2011 as shown in blue lines) and the days 
populated with several successive flare events (12-17 February, 2011 as shown in red lines). The power of 
oscillatory modes show enhancement in different frequency regimes on the days having 
flares as compared to the quiet day.}
\end{figure*}

In the following, we describe the analysis of GOLF and VIRGO/SPM observations for our 
investigation.

\subsection{Analysis of Velocity Observations from GOLF}
\label{sect:analyze_golf}
The 7-day long time series of velocity observations as obtained by the GOLF instrument is 
subjected to Fourier Transform to estimate the velocity power spectra for 
a time window of 8 hours for the different days as summarized in Table~1. 
In Figure~1, we show in the dashed black lines the 
Fourier Power Spectra obtained from these velocity oscillations. A comparison of these power spectra 
shows that various modes of oscillations in different 
frequency regimes are either getting enhanced or excited on the flare days with respect to the quiet day. 
We also apply a smoothing fit (Savitzky-Golay Fit; \citealp{Press1992}) as shown in the solid blue/red 
lines to these power spectra for estimating the power envelopes 
for the different days, which would be useful in proper comparison of the power spectra obtained 
for the flare days with that for the quiet day. It is to be noted that the Savitzky-Golay Fit 
(hereafter, S-G Fit) is a polynomial least-squares fit with weighted moving average, where 
weighting is given as a polynomial of an optimum degree depending on the characteristics of 
the data \citep{Press1992}. In Figure~2, we show the comparison of the 
Fourier Power Spectra for the flare days (shown in the solid red lines) with that obtained on 
the quiet day (shown in the solid blue lines). The comparative study of the power spectra as 
shown in the different panels of Figure~2 indicates that 
there is enhancement of the power of velocity oscillations on the flare days as 
compared to the quiet day.

On the other hand, it is well known that stochastic processes in the solar convection zone that excite 
the global modes of oscillations in the Sun \citep{Goldreich1994} could also experience 
changes on short time-scales thereby affecting the amplitudes of the oscillatory modes. 
Therefore, there could be a possibility that stochastic excitation of 
the global modes would have undergone a change over the period of our observations, 
viz., 11 February 2011 to 17 February 2011, thereby showing significant changes in the power of the 
oscillatory modes on the flare days as compared to the quiet day. In order to test this hypothesis, 
we have analyzed the disk-integrated velocity observations obtained by the GOLF instrument for 
8 hours of observations (08:00-16:00~UT) everyday during
the period from 19 April 2007 to 23 April 2007 during the extended solar magnetic-activity 
minimum \citep{Salabert2009}. This period 
during solar-activity minimum has no reported flares and thus can be used as another null-test period 
for our analysis. In Figure~3, we show the comparison of the smoothed (S-G Fit) Fourier 
Power Spectra obtained from the GOLF velocity observations for the aforementioned period. 
These power spectra do not show 
the extent of variations as seen in the case of flare days with respect to the quiet 
day (c.f.,~Figure~2). We further quantify the degree of relative variation in the power of 
global velocity oscillations over the aforementioned days during the solar minimum. It 
is found that the maximum relative variation in the integrated power as estimated from 
the non-smoothed power spectra of velocity oscillations over the days during the quiet 
period is about 9.85\% in the 2--5~mHz band ($p$ modes) and 
about 6.40\% in the 5--8~mHz band (high-frequency waves), which could be related to the 
stochastic variations in the power spectra due to the convective processes.

A similar analysis for the relative variation of power over the flare days with respect 
to the quiet days is also being presented at a later stage (Section~3.3) in this paper. 
There, we also compare the results obtained for the flare days with these results for the solar 
minimum phase.

\subsection{Analysis of Intensity Observations from VIRGO/SPM}
\label{sect:analyze_virgo}
We apply the same methodology used in GOLF to the VIRGO/SPM time series for the same 
time period. In Figure~4, we show the Fourier Power Spectra obtained from 
these intensity observations. These power spectra illustrate that various modes of 
oscillations at different 
frequencies are enhanced (or, excited) on the days with flares as compared to 
the ``null test'' (quiet day). In Figure~5, we show the comparison of 
the Fourier Power Spectra obtained for the flare days with that obtained on the 
quiet day. 
We observe from the comparison of power spectra shown in Figure~5 that 
there is enhancement of the power of intensity oscillations on the flare days as 
compared to the quiet day.

\subsection {Analysis of Variations in the Integrated Power on the Flare Days Relative to the Quiet Day}
\label{sect:analyze_power}

\begin{table*}
\centering
\caption{Analysis of relative variations in the integrated power in 2--5~mHz~band and 5--8~mHz~band for 
GOLF and VIRGO/SPM observations during the period from 11 February 2011 to 17 February 2011. A summary of
{\em GOES} soft X-ray flares (1--8~\AA~band) during the observing periods is also presented.}
\label{tab:GOLF_VIRGO_relative_power}
\begin{tabular}{llcccccccc} 
\hline
Date &
Flares &
\multicolumn{2}{c}{GOLF Integrated Power}  & 
\multicolumn{2}{c}{Rel. Variation in Power} & 
\multicolumn{2}{c}{VIRGO Integrated Power}  & 
\multicolumn{2}{c}{Rel. Variation in Power} \\
 & \multicolumn{2}{c}{(arbitrary units $*$ $10^{6}$)} & \multicolumn{2}{c}{(\%)} & \multicolumn{2}{c}{(arbitrary units $*$ $10^{5}$)} & \multicolumn{2}{c}{(\%)} &\\
& & (2--5~mHz) & (5--8~mHz) & (2--5~mHz) & (5--8~mHz) &
(2--5~mHz) & (5--8~mHz) & (2--5~mHz) & (5--8~mHz)\\
\hline
11 & 0 & 3.60 & 1.06 & - & - & 51.7 & 8.31 & - & - \\
12 & 3B, 1C & 5.26 & 1.26 & 46.11 & 18.87 & 63.70 & 9.78 & 23.21 & 17.69\\
13 & 2C, 1M & 4.54 & 1.21 & 26.11 & 14.15 & 63.40 & 9.57 & 22.63 & 15.16\\
14 & 1B, 4C, 1M & 5.25 & 1.24 & 45.83 & 16.98 & 62.43 & 9.52 & 20.75 & 14.56\\
15 & 2C, 1X & 4.50 & 1.20 & 25.00 & 13.20 & 81.20 & 10.07 & 57.05 & 28.76\\
16 & 4C, 1M & 4.30 & 1.18 & 19.44 & 11.32 & 59.22 & 9.30 & 14.50 & 11.91\\
17 & 6C & 5.81 & 1.30 & 61.38 & 22.64 & 78.53 & 9.96 & 52.28 & 19.86\\
\hline
\end{tabular}
Estimated stochastic variation in power: $\sim$~9.85\% in the 2--5~mHz~band, and $\sim$~6.40\% in the 5--8~mHz~band \\
\end{table*}

It is important to analyze and quantify the degree of variations in the power of the power spectra 
on the flare days relative to the quiet day in the frequency regimes of $p$ modes (2--5~mHz~band) 
and high-frequency waves (5--8~mHz~band) and to study the causal relation between the strength of the flares 
({\em GOES} soft X-ray peak flux) during the analyzing periods and the flare-induced 
enhancements seen in the power of global oscillation modes in the Sun. In Table~2, we 
present the analysis of integrated power obtained from the non-smoothed GOLF and VIRGO/SPM Fourier Power 
Spectra (c.f.,~Figures~1~and~3), respectively, for the period 11 February 2011 to 
17 February 2011. Our analysis shows the following results:

a. In the case of GOLF observations, we find that there is a variation in the integrated power of the 
global oscillation modes in the range of about 19\% to 61\% for the $p$-mode band and about 11\% to 22\% for 
the high-frequency band.

b. In the case of VIRGO/SPM observations, the variation in the integrated power of the global 
modes is in the range of about 14\% to 57\% for the $p$-mode band and about 
11\% to 28\% for the high-frequency band.

We also present a summary of {\em GOES} soft X-ray (1--8~\AA~band) flares during the same period 
for ready reference in Table~2.

The analysis of Table~2 illustrates that GOLF and VIRGO/SPM observations show minimum relative 
enhancement of integrated power in the $p$-mode band ($\sim$~19\% for GOLF and $\sim$~14\% for VIRGO) 
and in the high-frequency band ($\sim$~11\% for both, GOLF and VIRGO) on 16 February 2011, 
although 4 C-class and 1 M-class flares took place 
on this day during the observing period. On the other hand, the day 12 February 2011 with 
much lower strength of flares (3 B-class and 1 C-class flares) with respect to 16 February 2011, shows 
about 46\% and 18\% relative enhancements of integrated powers in $p$-mode band and high-frequency 
band, respectively, for GOLF observations and about 23\% and 17\% in the aforementioned two 
bands, respectively, for VIRGO observations. This is also illustrated in the comparison of the 
power envelopes of GOLF observations (c.f.,~Figure~2) and VIRGO observations (c.f.,~Figure~5). 
Similarly, a clear relation between the flare associated relative enhancements in the integrated 
power of the power spectra in the two frequency regimes (2--5~mHz~band, and 5--8~mHz~band) and 
the strength of the flares during the analyzing periods is also not seen for the other days 
considered in our analysis.

As it was shown in Section~3.1, the analysis of GOLF velocity observations for the period 
19 April 2007 to 23 April 2007 (solar minimum with no reported flares) reveals that normal 
stochastic excitation variations in the 
power of the power spectra over the aforementioned days is about 9.85\% 
in the $p$-mode band and 6.4\% in the high-frequency band. However, the flare associated minimum 
enhancements (19\% in the $p$-mode band and 11\% in the high-frequency band) observed in the 
analysis of GOLF data on 16 February 2011 are also reasonably higher relative to the maximum 
variation in power seen during the solar minimum. All other flare-days during 12 February 2011 to 
17 February 2011 are always showing much higher relative variation in power (for both, GOLF and VIRGO) 
with respect to the stochastic variations in the power spectra associated with convection.

\section{Discussion and Conclusions}
\label{sect:discussion}

We have analyzed the influence of flares on the global oscillation modes in the Sun using the 
Sun-as-a-star velocity and intensity observations obtained by the GOLF and 
VIRGO/SPM instruments, respectively. The chief findings of our investigation and the interpretations 
of results are as follows:

i. The Fourier Power Spectrum analysis of the GOLF and VIRGO observations indicates that there 
is enhancement in the power of predominant $p$ modes as well as 
high-frequency waves in the Sun during the flares, as compared to the quiet day. In this context, 
it is worthy mentioning that the GOLF instrument obtains velocity observations using the Na~I~D~lines 
which are formed in the upper solar photosphere, while the intensity data used in our analysis are 
obtained by VIRGO/SPM instrument at 862~nm, which is formed within 
the solar photosphere. Despite the fact that the two instruments sample different layers of the 
solar atmosphere using two different parameters (velocity v/s intensity), we have found that both 
these observations show the signatures of flare-induced global waves in the Sun. The GOLF and 
VIRGO/SPM observations have been used since the launch of {\em SOHO} to characterize the $p$-mode 
properties. The amplitudes of solar oscillations as detected by both instruments are more than two 
order of magnitudes above the photon noise level even at the Nyquist frequency (in the case of GOLF, 
refer to Figures~13~and~14 of \citealp{Garcia2005}). However, the limiting factor to measure $p$ modes 
is not the instrumental noise (e.g. photon noise), but solar noise composed of the sum of the 
convective background (mostly granulation at the $p$-mode frequencies) and the accumulated power 
of the unresolved high-degree modes (see for e.g., Figures~2~and~3 of \citealp{Garcia2013}).  
In the case of VIRGO/SPM, the Figure~3 of \citet{Garcia2013} illustrates that the 
amplitudes at $\nu_{max}$ are about two order of magnitudes above the noise level 
(assuming the instrumental noise level at the Nyquist frequency). However, considering only the 
amplitudes from the maximum of the modes against the region between the modes is $\sim$~1 order 
of magnitude ``above the noise''. The GOLF observations have a higher signal-to-noise ratio as the solar convective 
background is lower in velocity than intensity.

ii. It is observed that with flares the 
nature and degree of enhancements in the different oscillatory modes as seen in GOLF observations 
(c.f.,~Figure~2, and Table~2) and the VIRGO/SPM observations (c.f.,~Figure~5, and Table~2) 
are different, which could be assigned to the different responses to the flares by the two different 
observing layers in the solar atmosphere as sampled by these two instruments. The 
difference in the spectra of flares could also be the contributions from the different integration 
times being used in GOLF and VIRGO observations and the different observing parameters 
(velocity v/s intensity). \citet{Toutain1997} performed a correlation analysis between the VIRGO and 
GOLF observations for frequencies above 1~mHz ($p$~modes) for a 31-day long time series and found that 
the correlation of frequencies 
between the VIRGO/SPM (red channel at 862~nm) and GOLF observations is $\sim$~0.344 with a shift 
of maximum correlation by $\sim$~$+$273.69 s, which is not connected with the mean phase 
shift between intensity and velocity. This low correlation is apparently a combined effect of the 
different observing parameters at different wavelengths with different integration times of 
these two instruments. Accordingly, the power spectra with flares would show different 
behaviour.

iii. We observe dominant peaks in the spectra of flares around 4~mHz, which are seen on 
all the flare days for GOLF while on most of the flare days for VIRGO observations. This 
flare-induced enhanced power around 4~mHz has also been seen in local helioseismic wave power 
with a flare as analyzed by \citet{Gomez2012}. In general, it is also seen that there is 
enhanced power of oscillatory modes around 2~mHz in VIRGO observations on the flare days, 
which is not a commonly observed phenomenon. Hence, this indicates that flares have induced 
these oscillations.

iv. Our analysis shows that the flare associated changes seen in the power of the global modes of 
oscillations in the Sun in the $p$-mode band and the high-frequency band on the 
flare days as compared to the quiet day are always higher than the stochastic variations in the 
power spectra due to the convective processes. Hence, this relative enhancement in the 
power of global waves 
on the flare days with respect to the quiet day is the effect of flares or the different 
processes accompanying the flares taking place in the solar atmosphere.

v. It is noteworthy mentioning that various modes of oscillations in different frequency 
regimes are either getting 
enhanced or excited on the flare days with respect to the 
quiet day and there is a general increase in the power of global modes of oscillations in the 
Sun during the flares. 
However, we do not observe an explicit relation between the strength of the flares in the 
soft X-ray observations taking place 
during the analyzing periods for the flare days and the degree of enhancements in 
the power of oscillatory modes in the Sun as seen in the GOLF velocity and VIRGO/SPM intensity 
observations, respectively (c.f.,~Table~2). However, there is a possibility of any
flares happening on the far-side of the Sun during our analyzing periods and thereby 
contributing to the excitation of global waves in the Sun. Thus, 
our attempt to analyze the relation between the degree of enhancements seen in the power of 
global oscillation modes and the strength of flares as observed with {\em GOES} satellite suffers 
from the non-availability of flare observations on the far-side of the Sun.

vi. The aforementioned results also indicate that it is not only the flare energy that induces the 
global waves in the Sun, however, there are other factors influencing the power of these 
oscillatory modes during the flares. One of the probable scenario is that the fast global 
re-organization of the solar magnetic fields associated with the flares could lead to 
the impulsive changes in the Lorentz force vectors acting in the solar atmosphere. These 
impulsive changes in the Lorentz force are known to produce seismicity in the Sun 
(\citealp{Kumar2011, Kumar2016}, and references therein) thereby contributing to 
the enhancement of global modes of oscillations during the flares.

vii. The studies related to flare-induced local helioseismic wave power have provided 
some important clues, such as, (i) Not all the flares (or major flares) are associated with 
local seismic emissions in the Sun, (ii) In the case of flares which have been observed to 
induce local helioseismic emissions, only a fraction of the total flare energy is found to 
be converted into the acoustic emissions, and (iii) relatively smaller flares have sometimes 
shown larger effects of local helioseismic emissions (for a detailed discussion on these 
aspects, refer to the review paper by \citet{Donea2011}). Recently, \citet{Judge2014} 
showed that the energy of acoustic emissions in a sunquake was much more than the 
energy available from Lorentz force changes and radiative sources in an X1 flare in NOAA active 
region 12017 on 29 March 2014. Thus, there is also a lack of correlation between the flare-induced 
local helioseismic wave power and flare energies.

Altogether, these results could suffice our knowledge about the seismic responses to transient events, 
such as flares and the various impulsive processes accompanying the flares, taking place 
in the environment of the Sun. Also, it would be helpful in identifying the 
asteroseismic signatures of stellar flares using the knowledge gleaned from these solar results.

\section*{Acknowledgements}

We acknowledge the use of data from GOLF and VIRGO/SPM instruments on board {\em SOHO} space mission. 
The {\em SOHO} is a joint mission of NASA and ESA under co-operative agreement. We also acknowledge 
the use of data from {\em GOES-15} space mission of NASA. We are very much thankful to the referee 
for constructive comments and suggestions, which improved the presentation of our work in this 
paper. Thanks to Siddharath Sharma and Hemant Saini 
for doing initial projects on these observations. We are also thankful to P. Venkatakrishnan for 
useful discussions. BK acknowledges the support being provided by Udaipur Solar 
Observatory/Physical Research Laboratory and RAG acknowledges the support of CNES. SM would 
like to acknowledge support from NASA grants NNX15AF13G and NSF grant AST-1411685.




\bibliographystyle{mnras}
\bibliography{ms}
\bsp	
\label{lastpage}
\end{document}